\documentclass[draft]{agujournal2019}
\usepackage{url} 
\usepackage{lineno}
\usepackage[inline]{trackchanges} 
\usepackage{soul}
\usepackage[
    backend=biber,
    style=apa,
    uniquelist=false
  ]{biblatex}
\usepackage{dirtytalk}
\usepackage{amsthm}
\usepackage{amssymb}
\usepackage{amsmath}
\usepackage{mathtools}
\usepackage{xfrac}
\usepackage{gensymb}
\usepackage{graphicx}
\usepackage{multirow}
\usepackage{setspace}
\usepackage{caption}
\usepackage{makecell}
\newcommand{\defeq}{\vcentcolon=}

\usepackage{graphicx}
\usepackage{tikz}
\usetikzlibrary{shapes.geometric, arrows, calc}

\usepackage{xr}
\makeatletter
\newcommand*{\addFileDependency}[1]{
  \typeout{(#1)}
  \@addtofilelist{#1}
  \IfFileExists{#1}{}{\typeout{No file #1.}}
}
\makeatother

\newcommand*{\myexternaldocument}[1]{%
    \externaldocument{#1}%
    \addFileDependency{#1.tex}%
    \addFileDependency{#1.aux}%
}
\myexternaldocument{supplement_R1}

\draftfalse

\journalname{JGR: Solid Earth}

\begin{document}
\title{Embracing Data Incompleteness for Better Earthquake Forecasting}

%
%


\authors{L. Mizrahi\affil{1}\thanks{Sonneggstrasse 5, 8092 Zurich, Switzerland}, S. Nandan\affil{1}, and S. Wiemer\affil{1}}

\affiliation{1}{Swiss Seismological Service, ETH Zurich}


\correspondingauthor{Leila Mizrahi}{leila.mizrahi@sed.ethz.ch}




\begin{keypoints}
\item Two methods are proposed to invert ETAS parameters when catalog completeness varies with time.
\item Using pseudo-prospective experiments we compare the forecasting skill of our proposed models to a strong ETAS null model. 
\item Including small events in simulations yields increasingly superior forecasts for decreasing target magnitudes when using STAI correction.
\end{keypoints}

%
%

%
%


\begin{abstract}
    We propose two methods to calibrate the parameters of the epidemic-type aftershock sequence (ETAS) model based on expectation maximization (EM) while accounting for temporal variation of catalog completeness.
    The first method allows for model calibration on long-term earthquake catalogs with temporal variation of the completeness magnitude, $m_c$. 
    This calibration technique is beneficial for long-term probabilistic seismic hazard assessment (PSHA), which is often based on a mixture of instrumental and historical catalogs.
    The second method generalizes the concept of $m_c$, considering rate- and magnitude-dependent detection probability, and allows for self-consistent estimation of ETAS parameters and high-frequency detection incompleteness.
    With this approach, we aim to address the potential biases in parameter calibration due to short-term aftershock incompleteness, embracing incompleteness instead of avoiding it.
    Using synthetic tests, we show that both methods can accurately invert the parameters of simulated catalogs.
    We then use them to estimate ETAS parameters for California using the earthquake catalog since 1932.
    To explore how model calibration, inclusion of small events, and accounting for short-term incompleteness affect earthquakes' predictability, we systematically compare variants of ETAS models based on the second approach in pseudo-prospective forecasting experiments for California.
    Our proposed model significantly outperforms the ETAS null model, with decreasing information gain for increasing target magnitude threshold.
    We find that the ability to include small earthquakes for simulation of future scenarios is the primary driver of the improvement and that accounting for incompleteness is necessary.
    Our results have significant implications for our understanding of earthquake interaction mechanisms and the future of seismicity forecasting.
\end{abstract}

\section*{Plain Language Summary}
Our capability to detect earthquakes varies with time, on one hand because more and better instruments are being deployed over time, leading to long-term changes of detection capability.
On the other hand, earthquakes are more difficult to be detected when seismic activity is high, which manifests in short-term changes of detection capability.
Incomplete detection can lead to biases in epidemic-type aftershock sequence (ETAS) models used for earthquake forecasting.
We propose two methods that allow us to calibrate these models while accounting for long-term (first method) and short-term (second method) changes in detection capability, which allows us to use a larger and more representative fraction of the available data.
We test both methods on synthetic data and then apply them to the Californian earthquake catalog.
Using the second method, we test how small earthquakes can help us improve ETAS forecasts.
We find that the ability to include small earthquakes in simulations yields superior ETAS forecasts, and that it is necessary to correct for short-term incompleteness to achieve this superiority.
The positive effect on forecasting is strongest when forecasting relatively small events, and decreases when forecasting larger events.
These results have important implications for our understanding of earthquake interactions and for the future of earthquake forecasting.

\section*{Keywords}
data incompleteness, model inversion, ETAS, earthquake forecasting

\section{Introduction}
    One of the key challenges in the field of statistical seismology, or seismology in general, is the development of accurate earthquake forecasting models, with epidemic-type aftershock sequence (ETAS) models (see \cite{ogata1998space}; \cite{veen2008estimation}; \cite{nandan2017objective}) currently being widely used for this purpose.
    ETAS models account for the spatio-temporal clustering of earthquakes, and they have been shown in retrospective, pseudo-prospective, and prospective forecasting experiments to be among the best-performing earthquake forecasting models available today (\cite{nandan2019forecastingrates}; \cite{taroni2018prospective}, \cite{woessner2011retrospective}; \cite{cattania2018forecasting}; \cite{mancini2019improving}; \cite{mancini2020predictive}).
    Furthermore, they are used for operational earthquake forecasts by the USGS (\cite{field2017synoptic}), in Italy (\cite{marzocchi2014establishment}), and New Zealand (\cite{rhoades2016retrospective}).
    
    A fundamental requirement for reliable parameter estimation of the ETAS model is the completeness of the training catalog above the magnitude of completeness, $m_c$. 
    As we can impossibly know with certainty what we did not observe, $m_c$ itself needs to be estimated, and numerous approaches to this problem have been proposed (\cite{wiemer2000minimum}; \cite{cao2002temporal}; \cite{woessner2005assessing}; \cite{amorese2007applying}; \cite{rydelek1989testing}; see \textcite{mignan2012estimating} for an overview). 
    $m_c$ is known to vary in space and time, due to gradual improvement of the seismic network, software upgrades, and so on. 
    Several methods have been proposed to estimate its spatial and temporal variation (\cite{wiemer2000minimum}; \cite{mignan2011bayesian};\cite{woessner2005assessing}; \cite{amato2008performance}; \cite{nanjo2010analysis}; \cite{hutton2010earthquake}; \cite{mignan2014fifty}; \cite{schorlemmer2008probability}; \cite{hainzl2016rate}). 
    In particular, \textcite{hainzl2016rate} addresses an additional important cause of variation in time of $m_c$, short-term aftershock incompleteness (STAI). 
    Because earthquakes strongly cluster in time, seismic networks can only capture a subset of events during periods of high activity (\cite{kagan2004short}). 
    
    As mentioned earlier, a reliable estimation of ETAS parameter depends on a reliable estimate of $m_c$.
    Although the biasing effects on ETAS parameter estimates caused by data incompleteness are known and discussed (\cite{hainzl2016rate}; \cite{seif2017estimating}; \cite{zhuang2017data}), nearly all applications of the ETAS model assume for simplicity a global magnitude of completeness for the entirety of the training period.
    This assumption is problematic in several ways. 
    
    First, in order to be complete for the entire training period, the modeller is often forced to use very conservative estimates of $m_c$, as a result completely ignoring abundant and high-quality data from more complete periods. 
    Furthermore, $m_c$ is often assumed to be equal to the minimum magnitude of earthquakes that can trigger aftershocks, $m_0$, and this conservative assumption can introduce a bias to ETAS parameter estimates. 
    This idea that earthquakes below $m_c$ are relevant for our understanding of earthquakes' clustering behavior was thoroughly discussed by \textcite{sornette2005constraints}, who pointed out the important distinction between $m_0$ and $m_c$, also providing constraints for $m_0$. 
    Although small earthquakes trigger fewer aftershocks than large ones do, \textcite{marsan2005role}, as well as \textcite{helmstetter2005importance}, found that small earthquakes, being more numerous, are as important as large ones for earthquake triggering.
    Thus, it is natural to assume that a larger difference between $m_c$ and $m_0$ will lead to a larger bias in the estimated parameters.
    
    Alternatively, one may estimate ETAS parameters from catalogs with restricted space-time volume which can have low overall $m_c$ values.
    Parameters estimated in this way can however be dominated by one or two sequences and may not represent long-term behavior, thus making the use of ETAS models non-ideal for long-term probabilistic seismic hazard assessment (PSHA). 
    Instead, the modellers rely on smoothed seismicity approaches based on declustered catalogs (see e.g. \cite{gerstenberger2020probabilistic}; \cite{petersen20182018}; \cite{wiemer2009probabilistic}), which is a problematic approach due to the biasing effects of declustering on the size distribution of mainshocks, and thus on the estimated seismic hazard (\cite{mizrahi2021effect}). In this regard, \textcite{marzocchi2014some} discussed the need for spatial declustering so as not to distort future seismic hazard, and \textcite{llenos2020regionally} proposed an approach to calculate regionally optimized background earthquake rates from ETAS to be used for the U.S. Geological Survey National Seismic Hazard Model (NSHM), stressing the need for methods to address catalog heterogeneities such as time-dependent incompleteness.
    
    Additionally, with the assumption of a constant overall $m_c$, the crucial requirement of completeness of the training catalog is not fulfilled during large aftershock sequences, which can bias the estimated parameters.
    Several studies have highlighted the importance of considering short-term variation of $m_c$ in the context of ETAS models.
    These include \textcite{hainzl2016rate} and \textcite{hainzl2016apparent}, who modeled STAI based on the short-term rate of earthquakes, bringing into relation true and apparent triggering laws; \textcite{stallone2020missing}, who proposed a method to stochastically replenish catalogs suffering from STAI, to be used for better operational earthquake forecasting and hazard assessment, albeit without addressing the effectiveness of the method in this regard; \textcite{zhuang2017data}, who showed that estimating ETAS parameters using a replenished catalog is more stable with respect to cutoff magnitude; \textcite{omi2014estimating}, who proposed a method to estimate parameters of the ETAS model from incompletely observed aftershock sequences, by statistically modelling detection deficiency.
    
    In this article, we thoroughly address the use of small earthquakes for seismic hazard forecasting.
    For this, we develop two complementary methods with which long-term (first method) and short-term (second method) temporal variations of $m_c$ can be accounted for when calibrating ETAS models and when issuing ETAS-based forecasts.
    The first method extends the expectation maximization scheme for ETAS parameter inversion (\cite{veen2008estimation}) for application to training catalogs with time-varying completeness magnitude $m_c(t)$.
    This simultaneously allows the inclusion of historical data in the parameter inversion, as well as the inclusion of small magnitude events, which make up a large fraction of data and can enable the ability to more clearly illuminate faults.
    ETAS models can hence be trained on a more representative and informative set of data, which in some areas facilitates a more appropriate approach to PSHA.
    With the second method proposed in this article, we want to utilize the knowledge about clustering derived using the ETAS model to quantitatively estimate the level of completeness of a catalog at any given time, and then use this knowledge to minimize the incompleteness-induced bias in the ETAS model. 
    We approach this issue by generalizing the notion of $m_c$, moving from a binary completeness space (complete versus incomplete) to a continuous-valued completeness space by means of a magnitude-dependent detection probability $-$ embracing incompleteness instead of avoiding it, as has been proposed previously by \textcite{ogata2006immediate} and \textcite{omi2014estimating}.
    While the first method described in this article allows $m_c(t)$ as an input to the ETAS parameter calibration, which makes it powerful in a long-term context, the second method addresses the additional challenge of estimating short-term variations of completeness.
    To understand their abilities and limitations, we subject both methods to rigorous synthetic tests.
    Then, we apply them to Californian earthquake data and interpret the results in light of the findings of the synthetic tests.
    Using the second approach, we systematically assess how the inclusion of small earthquakes, which may be incompletely detected, affects the performance of earthquake forecasts.
    We conduct pseudo-prospective 30-day forecasting experiments for California, designed to answer several questions:
    Does our new model outperform the current state of the art?
    If so, what is the role of the newly estimated ETAS parameters in this improvement?
    Similarly, what is the role of newly included small earthquakes in this improvement, and the role of the estimated high-frequency detection incompleteness?
    How do the models perform for different target magnitude thresholds?
    
    The remainder of the paper is structured as follows.
    Section \ref{sec:data} describes the earthquake catalog that was used in this analysis. 
    The modified ETAS parameter inversion methods are presented in Section \ref{sec:zeroone_etas} for time-varying $m_c$, and in Section \ref{sec:pdet_etas} for time-varying probabilistic detection incompleteness.
    Sections \ref{sec:detprob} and \ref{sec:petai} describe the formulation of probabilistic detection incompleteness and the algorithm for joint estimation of ETAS parameters and detection probability. 
    Section \ref{sec:synth} presents synthetic tests for both methods,
    and Section \ref{sec:applied} presents applications of both methods to the Californian data.
    Section \ref{sec:pseudo} describes pseudo-prospective forecasting experiments used to assess the impact of the newly acquired information on the forecastability of earthquakes in California. 
    Finally, in Section \ref{sec:conclusion}, we present our conclusions.

\section{Data}
    \label{sec:data}
    In this article, we use the ANSS Comprehensive Earthquake Catalog (ComCat) provided by the U.S. Geological Survey.
    We adopt the preferred magnitudes as defined in ComCat, and use as study region the collection area around the state of California as proposed in the RELM testing center (\cite{schorlemmer2007relm}).
    We consider events of magnitude $M\geq0.0$, with magnitudes rounded into bins of size $\Delta M = 0.1$.
    For the major part of the study, the time frame used is January 1, 1970 until December 31, 2019. For the analysis of long-term variations in $m_c$, we extend the time frame to start on January 1, 1932, when instrumentation was introduced to the Californian seismic network (\cite{felzer2007appendix}).
    
    Whenever ETAS parameters are inverted, we use the first fifteen years of data to serve as auxiliary data. Thus, the start of the primary catalog is either January 1985, or January 1947. Earthquakes in the auxiliary catalog may act as triggering earthquakes in the ETAS model, but not as aftershocks.
    
    To estimate a constant magnitude of completeness of the catalog, we use the method described by \textcite{mizrahi2021effect} with an acceptance threshold value of $p=0.1$, which yields $m_c = 3.1$ for the time period between 1970 and 2019. 
    This method is adapted from \textcite{clauset2009power} and jointly estimates $m_c$ and the $b$-value of the Gutenberg-Richter law (\cite{gutenberg1944frequency}) describing earthquake size distribution.
    It compares the Kolmogorov-Smirnov (KS) distance between the observed cumulative distribution function (CDF) and the fitted GR law to KS distances obtained for magnitude samples simulated from said GR law.
    A value of $m_c$ is accepted if at least a fraction of $p=0.1$ of KS distances is larger than the observed one.
    
\section{Model}

    \subsection{ETAS parameter inversion for time-varying $m_c$}
        \label{sec:zeroone_etas}
        Consider an earthquake catalog
        
        \begin{equation}
            C=\{e_i=(m_i, t_i, x_i, y_i), i\in\{1, \dots, n\}\}
        \end{equation}
        consisting of events $e_i$ of magnitudes $m_i$ which occur at times $t_i$ and locations $(x_i, y_i)$.
        Furthermore, consider a time-varying magnitude of completeness $m_c(t)$ defined for all $t_i$.
        We say that the catalog is complete if $m_i \geq m_c(t_i) \forall i$.
        
        The ETAS model describes earthquake rate as
        
        \begin{equation}
        \label{eq:etasrate}
            l(t,x,y) = \mu + \sum_{i: t_i < t} g(m_i, t-t_i, x-x_i, y-y_i).
        \end{equation}
        That is, the sum of background rate $\mu$ and the rate of all aftershocks of previous events $e_i$.
        The aftershock triggering rate $g(m, \Delta t, \Delta x, \Delta y)$ describes the rate of aftershocks triggered by an event of magnitude $m$, at a time delay of $\Delta t$ and a spatial distance $(\Delta x, \Delta y)$ from the triggering event.
        We here use the definition
        
        \begin{equation}
            \label{eq:g_etas}
            g(m,\Delta t, \Delta x,\Delta y) = \frac{k_0 \cdot e^{a(m-m_{ref})}}{\frac{(\Delta t+c)^{1+\omega}}{e^{\frac{-\Delta t}{\tau}}}\cdot \left((\Delta x^2 + \Delta y^2) + d\cdot e^{\gamma (m-m_{ref})}\right)^{1+\rho}},
        \end{equation}
        as in \textcite{nandan2017objective}.
        
        To calibrate the ETAS model, the nine parameters to be optimized are the background rate $\mu$ and the parameters $k_0, a, c, \omega, \tau, d, \gamma, \rho$ which parameterize the aftershock triggering rate $g(m, t, x, y)$ given in Equation \ref{eq:g_etas}. Implicitly, the model assumes that all earthquakes with magnitude larger than or equal to $m_{ref}$ can trigger aftershocks.
        We build on the expectation maximization (EM) algorithm to estimate the ETAS parameters (\cite{veen2008estimation}). In this algorithm, the expected number of background events $\hat n$ and the expected number of directly triggered aftershocks $\hat l_i$ of each event $e_i$ are estimated in the expectation step (E step), along with the probabilities $p_{ij}$ that event $e_j$ was triggered by event $e_i$, and the probability $p_{j}^{ind}$ that event $e_j$ is independent. Following the E step, the nine parameters are optimized to maximize the complete data log likelihood in the maximization step (M step). 
        E and M step are repeated until convergence of the parameters.
        The usual formulation of the EM algorithm defines
            
        \begin{align}
            p_{ij} &= \frac{g_{ij}}{\mu + \sum_{k: t_k <t_j} g_{kj}},\label{eq:pij}\\
            p_{j}^{ind} &= \frac{\mu}{\mu + \sum_{k: t_k <t_j} g_{kj}} \label{eq:p_ind},
        \end{align}
        with $g_{kj} = g(m_k, t_j-t_k, x_j-x_k, y_j-y_k)$ being the aftershock triggering rate of $e_k$ at location and time of event $e_j$. 
        For a given target event $e_j$, Equations (\ref{eq:pij}-\ref{eq:p_ind}) define $p_{ij}$ to be proportional to the aftershock occurrence rate $g_{ij}$, and $p_j^{ind}$ to be proportional to the background rate $\mu$.
        As an event must be either independent or triggered by a previous event, the normalization factor 
        $\Lambda_j \defeq \mu + \sum_{k: t_k <t_j} g_{kj}$
        in the denominator of Equations \ref{eq:pij}-\ref{eq:p_ind} stipulates that $p_j^{ind} + \sum_{k: t_k <t_j} p_{kj} = 1$.
        This relies on the assumption that all potential triggering earthquakes of $e_j$ were observed, that is, all events prior to $t_j$ above the reference magnitude (minimum considered magnitude), $m_{ref}$ were observed.
        To fulfill this requirement, most applications of the method define $m_{ref}$ to be equal to the constant value of $m_c$.
        
        For the case of time-varying $m_c(t)$, we define $m_{ref}\defeq \min_i\{m_c(t_i)\}$, the minimum $m_c(t_i)$ for times $t_i$ of events in the complete catalog.
        This implies that for the times when $m_c(t)>m_{ref}$ the requirement of complete recording of all potential triggers may be violated. 
        Events whose magnitudes fall between $m_{ref}$ and $m_c(t)$ are not part of the complete catalog and are considered to be unobserved (even though they may have been detected by the network).
        Hence, the normalization factor $\Lambda_j$ (the denominator of Equations \ref{eq:pij}-\ref{eq:p_ind}) needs to be adapted to account for the possibility that $e_j$ was triggered by an unobserved event.
        
        Consider
        \begin{equation}
            \label{eq:xi_def}
            \xi(t) = \frac{
                \int_{m_{ref}}^{m_c(t)} f_{GR}(m) \cdot G(m) \,dm
            }{
                \int_{m_c(t)}^{\infty} f_{GR}(m) \cdot G(m) \,dm
            },
        \end{equation}
        the ratio between the expected number of events triggered by an unobserved event and the expected number of events triggered by an observed event at time $t$. 
        Here, $f_{GR} = \beta \cdot e^{-\beta \cdot (m - m_{ref})}$ is the probability density function of magnitudes according to the GR law, 
        and $G(m) = \int_0^{\infty}\iint_{R} g(m, t, x, y) \,dx\,dy\,dt$ is the total number of expected aftershocks larger than $m_{ref}$ of an event of magnitude $m$.
        Note that in the calculation of $G(m)$ we make the simplifying assumption that the considered region $R$ extends infinitely in all directions, allowing a facilitated, asymptotically unbiased estimation of ETAS parameters (\cite{schoenberg2013facilitated}).
        Analogously,
        
        \begin{equation}
            \label{eq:zeta_def}
            \zeta(t) = \frac{
                \int_{m_{ref}}^{m_c(t)} f_{GR}(m)\,dm
            }{
                \int_{m_c(t)}^{\infty} f_{GR}(m)\,dm
            }
        \end{equation}
        is the ratio between the expected fraction of unobserved events and the expected fraction of observed events at time $t$.
        If $\beta > a-\rho\gamma$, both $\xi(t)$ and $\zeta(t)$ are well-defined and we have that
        
        \begin{align}
            \xi(t) &= e^{- (a-\beta-\rho\gamma)\cdot\Delta m(t)} - 1,\\
            \zeta(t) &= e^{\beta \cdot \Delta m(t)} - 1,
        \end{align}
        where $\Delta m(t) = m_c(t) - m_{ref}$.
        Consider the productivity exponent $\alpha \defeq a - \rho\cdot \gamma$, which describes the exponential relationship between aftershock productivity and magnitude of an event.
        The condition that $\beta$ is larger than the productivity exponent $\alpha$  is generally fulfilled in naturally observed catalogs (\cite{helmstetter2003earthquake}).
        If this were not the case, earthquake triggering would be dominated by large events and one would need to introduce a maximum possible magnitude for both denominators to be finite (see available equations in \textcite{sornette2005apparent}; \textcite{sornette2005constraints}).
        The normalization factor $\Lambda_j$ consists of the sum of background rate and aftershock rates of all events which happened prior to $e_j$.
        In the case of time-varying $m_c$, besides the possibilities of being a background event or being triggered by an observed event, the event $e_j$ can also be triggered by an unobserved event.
        We thus generalize $\Lambda_j$ by adding to the rate of aftershocks $g_{kj}$ of each observed triggering event $e_k$ the expected rate of aftershocks of unobserved triggering events at that time, $g_{kj}\cdot \xi(t_k)$.
        This yields $\Lambda_j = \mu + \sum_{k: t_k <t_j} g_{kj} \cdot (1 + \xi(t_k))$ and thus the generalized definition of $p_{ij}$ and $p_j^{ind}$ is given by
        
        \begin{align}
            \label{eq:pij_new}
            p_{ij} &= \frac{g_{ij}}{\mu + \sum_{k: t_k <t_j} g_{kj} \cdot (1 + \xi(t_k))} ,\\
            p_{j}^{ind} &= \frac{\mu}{\mu + \sum_{k: t_k <t_j} g_{kj} \cdot (1 + \xi(t_k))}.
        \end{align}
            
        Note that the probability $p_{uj}$ that event $e_j$ was triggered by an unobserved event is given such that $p_j^{ind} + p_{uj} + \sum_{k: t_k <t_j} p_{kj} = 1$. 
        In the above equations, the special case of $m_c(t)\equiv m_{ref}$ is accounted for when $\xi(t)\equiv 0$.
        In this special case, $\hat n$ and $\hat l_i$ are obtained by summing independence probabilities ($\hat n = \sum_j p_j^{ind}$) and triggering probabilities ($\hat l_i = \sum_j p_{ij}$), respectively.
        In the generalized case however, $\hat n$ and $\hat l_i$ are the estimated number of background events and aftershocks of event $e_i$ above $m_{ref}$, which includes unobserved events.
        Similarly to inflating the triggering power, we hence inflate the observed event numbers to account for unobserved events.
        Whenever an event is observed at time $t_j$, we expect that $\zeta(t_j)$ events occurred under similar circumstances (with same independence and triggering probabilities), but were not observed.
        This yields
        
        \begin{align}
            \hat n &= \sum_j p_j^{ind}\cdot (1 + \zeta(t_j)),\\
            \hat l_i &= \sum_j p_{ij} \cdot (1 + \zeta(t_j)).\label{eq:lhat_new}
        \end{align}
        With these adapted definitions of $p_{ij}, p_j^{ind}, \hat{n}$ and $\hat{l_i}$ (Equations \ref{eq:pij_new} - \ref{eq:lhat_new}), ETAS parameters can be inverted using the procedure described by \textcite{veen2008estimation}.

    \subsection{ETAS parameter inversion for time-varying probabilistic detection}
        \label{sec:pdet_etas}
        To overcome the binary view of completeness which forces us to disregard earthquakes which were detected but happen to fall between $m_{ref}$ and $m_c(t)$, we can take the generalization of the EM algorithm for ETAS parameter inversion one step further by introducing a time and magnitude-dependent probability of detection,
        
        \begin{align*}
            f\colon \mathbb{R}_{\ge m_{ref}}\times \mathbb{R}&\longrightarrow [0, 1]\\
            (m, t) &\mapsto p.
        \end{align*}
        To be able to account for such a probabilistic concept of catalog completeness in the ETAS inversion algorithm, one needs to generalize $\xi(t)$ and $\zeta(t)$ (Equations \ref{eq:xi_def} and \ref{eq:zeta_def}).
        In contrast to before, the magnitude of an event does not determine whether or not the event has been detected.
        We therefore adapt the bounds of integration in numerator and denominator such that all events above magnitude $m_{ref}$ are considered.
        To obtain the expected number of earthquakes triggered by observed and unobserved events, the integrands are multiplied by the probability of the triggering events to be observed, $f(m, t)$, or unobserved, $(1-f(m, t))$, respectively.
        The generalized formulations of $\xi(t)$ and $\zeta(t)$ then read
        
        \begin{equation}
                \xi(t) = \frac{
                    \int_{m_{ref}}^{\infty} (1 - f(m, t)) \cdot f_{GR}(m) \cdot G(m) \,dm
                }{
                    \int_{m_{ref}}^{\infty} f(m, t) \cdot f_{GR}(m) \cdot G(m) \,dm
                },\label{eq:xi_final}
        \end{equation}
        and
        
        \begin{equation}
            \zeta(t) = \frac{
                \int_{m_{ref}}^{\infty} (1 - f(m, t)) \cdot f_{GR}(m)\,dm
            }{
                \int_{m_{ref}}^{\infty} f(m, t) \cdot f_{GR}(m)\,dm
            }.\label{eq:zeta_final}
        \end{equation}
        For compatible choices of $f(m, t)$, $f_{GR}(m)$, $G(m)$, we find that $\xi(t)$ and $\zeta(t)$ are well-defined.
        Consider for instance the special case of binary detection, where $f(m, t)$ is defined via the Heaviside step function $H$ as  $f_{bin}(m, t) = H(m - m_c(t))$, which is equal to $1$ if $m\ge m_c(t)$ and $0$ otherwise.
        This is the case discussed in the previous section, for which we have well-definedness if $\beta > a - \rho\gamma$.
        
        The reference magnitude $m_{ref}$ is a model constant. Smaller values of $m_{ref}$ allow the modeller to use a larger fraction of the observed catalog, which can be especially useful in regions with less seismic activity.
        
        Note that both generalizations of the ETAS inversion algorithm (for time-varying completeness or for time-varying probabilistic detection) can without further modification be applied when $m_c$ or detection probability vary with space.
        The formulation is based on the assumption that the behavior of observed events is locally representative (in space and/or time) of the behavior of unobserved events. 
    
    \subsection{Rate-dependent probabilistic detection incompleteness}
        
        \label{sec:detprob}
        In this section we present our approach to define $f(m, t)$, where the temporal component is purely driven by the current rate of events.
        Note that this means we only capture changes in detection due to changes in short-term circumstances, and neglect long-term changes due to network updates.
        We make the following simplifying assumptions.
        \begin{itemize}
            \item Any earthquake will obstruct the entire seismic network from detecting smaller earthquakes for a duration of $t_R$ (recovery time of the network). 
            \item Magnitudes of events that are simultaneously blocking the network are distributed according to the time-invariant Gutenberg-Richter law which also describes the magnitude distribution of the full catalog (\cite{gutenberg1944frequency}).
        \end{itemize}
        \textcite{de2018overlap} found that short-term aftershock incompleteness can be well explained in terms of overlapping seismic records, while instrumental coverage of an area plays a subsidiary role.
        Nevertheless, assuming $t_R$ to be independent of the magnitude of the event, and independent of the spatial distance between the event and the locations of interest, is certainly a major simplification which could be refined in subsequent studies.
        \\
        
        Conveniently, the ETAS model provides a simple way of calculating the current rate of events in the region $R$ as
        
        \begin{equation}
        \label{eq:currentrate}
            \lambda(t) = \iint_{R} l(t, x, y) \,dx \,dy = \iint_{R} \mu + \sum_{i: t_i < t}  g(m_i, t-t_i, x-x_i, y-y_i) \,dx\,dy.
        \end{equation}
        For the remainder of this paper, we will refer to the current rate of events in the region, $\lambda(t)$, simply as the current rate of events.
        The probability $f(m, t)$ of an earthquake to be detected is then given by the probability of it being the largest of all the earthquakes that are currently blocking the network. Consider
        
        \begin{equation}
            \label{eq:defprob}
            f(m, t) = \left(1 - e^{-\beta\cdot(m - m_{ref})}\right)^{t_R \cdot \lambda(t)}.
        \end{equation}
        Here, $t_R \cdot \lambda(t)$ is an approximation of the expected number of events blocking the network at time $t$, and the term $1 - e^{-\beta\cdot(m - m_{ref})}$ is the probability of any given earthquake's magnitude falling between $m_{ref}$ and $m$, where $\beta = b \cdot \ln{10}$ is the exponent in the GR law with basis $e$.
        Thus, $f(m, t)$ is the probability that in the set of $t_R \cdot \lambda(t)$ events currently blocking the network, all of them have a magnitude of less than $m$, which is the condition for an event of magnitude $m$ to be detected.
        Because the time-dependence of $f(m, t)$ is solely controlled by the time-dependence of $\lambda$, we here use the terms $f(m, t)$ and $f(m, \lambda)$ interchangeably.
        
        Plugging this definition of $f(m, t)$ into Equations \ref{eq:xi_final} and \ref{eq:zeta_final}, and setting $\kappa \defeq -\frac{a-\rho\gamma}{\beta}$, we obtain
        
        \begin{align}
            \xi(t) &= \frac{1}{(\kappa + 1)\cdot \rm B(\kappa + 1, t_R\cdot \lambda(t) + 1)} -1,\label{eq:xi}\\
            \zeta(t) &= t_R\cdot \lambda(t)\label{eq:zeta},
        \end{align}
        so long as $\beta > a - \rho\gamma$, where $\rm B$ is the Beta function. A positive background rate $\mu>0$ ensures $\lambda(t)>0\quad \forall t$. Expressions analogous to (\ref{eq:xi}) and (\ref{eq:zeta}) hold when alternative exponents are chosen instead of $t_R\cdot \lambda$ in the definition of $f(m,\lambda)$ (Equation \ref{eq:defprob}).\\
        
        The network recovery time $t_R$ and the current event rate $\lambda(t)$ at the times $t_i$ of all earthquakes $e_i$ need to be estimated from the data.
        
    \subsection{Estimating probabilistic epidemic-type aftershock incompleteness (PETAI)}
        \label{sec:petai}
        \subsubsection{Estimation of $(t_R, \beta)$ when $\lambda_i$ are known}
        \label{sec:tr_est}
        The function $f(m, t)$ brings with it two parameters, $t_R$ and $\beta$, which need to be estimated in addition to the ETAS parameters.
        We here describe how $t_R$ and $\beta$ can be jointly estimated using a maximum likelihood approach for the case when current event rates $\lambda_i = \lambda(t_i)$ are known. 
        In reality, the $\lambda_i$ have to be estimated themselves.
        This is described in Section \ref{sec:lambda_given_rest}.
        
        In the case when the true ETAS parameters, as well as the current event rates $\lambda(t_i)$ for all events $e_i$ in the primary catalog $\{e_1, \dots, e_n\}$, are known, the GR-law exponent $\beta$ and the network recovery time $t_R$ can be estimated by optimizing the log-likelihood $\mathcal{LL}$ of observing the catalog at hand.
        
        \begin{align}
            \mathcal{LL} =& \sum_{i=1}^{n} \left(\ln{(\nu_i + 1) - \ln{N}}\right)\nonumber\\
            &+\sum_{i=1}^{n} \left(\nu_i \cdot \ln{(1 - e^{-\beta \cdot (m_i - m_{ref})})}\right)\label{eq:normdet}\\
            &+\sum_{i=1}^{n} \left(\ln{\beta} - \beta\cdot(m_i - m_{ref})\right),\nonumber
        \end{align}
        where $N=\sum_{i=1}^{n} (\nu_i +1)$, and $\nu_i = t_R \cdot \lambda(t_i)$ is an approximation of the expected  number  of  events  blocking  the  network  at  time $t_i$. The expression for $\mathcal{LL}$ given above is valid in general for alternative exponents $\nu_i$ in the definition of detection probability (Equation \ref{eq:defprob}).
        $\mathcal{LL}$ is derived from the likelihood $\mathcal{L}_i$ of an event to have magnitude $m_i$ and to be observed during a current event rate of $\lambda_i = \lambda(t_i)$, and the current event rate being $\lambda_i$,
        
        \begin{equation}
            \mathcal{L}_i = 
                f_{emp}(\lambda_i) \cdot f_{GR}(m_i) \cdot f_{det}(m_i, \lambda_i),
        \end{equation}
        where $f_{GR}(m)$ is the probability density function of magnitudes given by the GR law, $f_{det}(m, \lambda)$ is the detection probability as defined in Equation \ref{eq:defprob}, and 
        
        \begin{equation}
            f_{emp}(\lambda) = 
            \begin{cases}
                \frac{t_R\cdot\lambda + 1}{\sum_i(t_R\cdot\lambda_i + 1)},& \text{if } \lambda\in\{\lambda_1, \dots, \lambda_n\}\\
                0,              & \text{otherwise}
            \end{cases}
        \end{equation}
        is the empirical density function of event rates.
        $f_{emp}(\lambda)$ is defined such that
        
        \begin{equation}
            \sum_{i=1}^{n} f_{emp}(\lambda_i) = 1
        \end{equation}
        and
        \begin{equation}
            \label{eq:empprop}
            f_{emp}(\lambda_i) \propto \frac{1}{
                \int_{m_{ref}}^{\infty} f_{GR}(m) \cdot f_{det}(m, \lambda_i)\,dm
            }
            = \lambda_i\cdot t_R + 1, \quad \forall i = 1, \dots, n.
        \end{equation}
        Without the latter condition (Equation \ref{eq:empprop}), we would wrongly assume that the values $\lambda(t_i)$ were uniformly drawn from the true distribution of event rates.
        However, in our sample of $\lambda_i$, large values of $\lambda$ are underrepresented, because during times $t$ when $\lambda(t)$ is high, events are less likely to be detected, and those times and their corresponding rates are thus less likely to be part of our sample.
        Defining $f_{emp}(\lambda_i)$ to be inversely proportional to the fraction of events that are observed when the current rate is $\lambda_i$ corrects for this under-representation.
        This yields
        
        \begin{equation}
            \mathcal{L}_i = \frac{\nu_i + 1}{\sum_j(\nu_j + 1)} \cdot \beta \cdot e^{-\beta\cdot(m_i - m_{ref})} \cdot \left(
                1 - e^{-\beta\cdot(m_i-m_{ref})}
                \right)^{\nu_i},
        \end{equation}
        which explains the term for $\mathcal{LL}$ (Equation \ref{eq:normdet}). Figure S1 shows the log likelihood of a synthetic test catalog for different values of $t_R$ and $\beta$ when $\lambda_i$ are known. The resulting estimators match the data-generating parameters.

        \subsubsection{Estimation of $\lambda_i$ when ETAS parameters and $(t_R, \beta)$ are known}
        \label{sec:lambda_given_rest}
        On one hand, the $\lambda_i$ depend on the ETAS parameters (see Equation \ref{eq:currentrate}). 
        On the other hand, the sum of aftershocks of previous earthquakes in the definition of $\lambda(t)$ (Equation \ref{eq:currentrate}) does not account for aftershocks of events that were not detected.
        As in the ETAS parameter inversion, to account for aftershocks of undetected events in the calculation of $\lambda(t)$, we inflate the triggering power of each event $e_i$ by a factor of $1 + \xi(t_i)$ and define
        
        \begin{align}
            \label{eq:lambda_new}
        \lambda(t) = \iint_{R} \mu \,dx\,dy + \sum_{i: t_i < t}(1 + \xi(t_i)) \cdot \iint_{R} g(m_i, t-t_i, x-x_i, y-y_i) \,dx\,dy.
        \end{align}
        
        \subsubsection{Estimation of $\lambda_i$ and $(t_R, \beta)$ when ETAS parameters are known}
        \label{sec:mathcal_I}
        $\xi(t)$ however requires knowledge of $(t_R, \beta)$ (see Equation \ref{eq:xi}).
        This implies that even when ETAS parameters are fixed, an additional, lower-level circular dependency dictates the relationship between $(\lambda_i)_{i=1,\dots,n}$ and $(t_R, \beta)$.

        To fully estimate the high-frequency probabilistic detection incompleteness, given fixed ETAS parameters, we recursively re-estimate $(\lambda_i)_{i=1,\dots,n}$ (see Section \ref{sec:lambda_given_rest}) and $(t_R, \beta)$ (see Section \ref{sec:tr_est}), until $(t_R, \beta)$ meets a convergence criterion, starting with an informed or random initial guess for $(t_R, \beta)$. 
        
    \subsection{PETAI inversion algorithm} 
\begin{figure}[ht]
        \centering
        \includegraphics[width=0.5\textwidth]{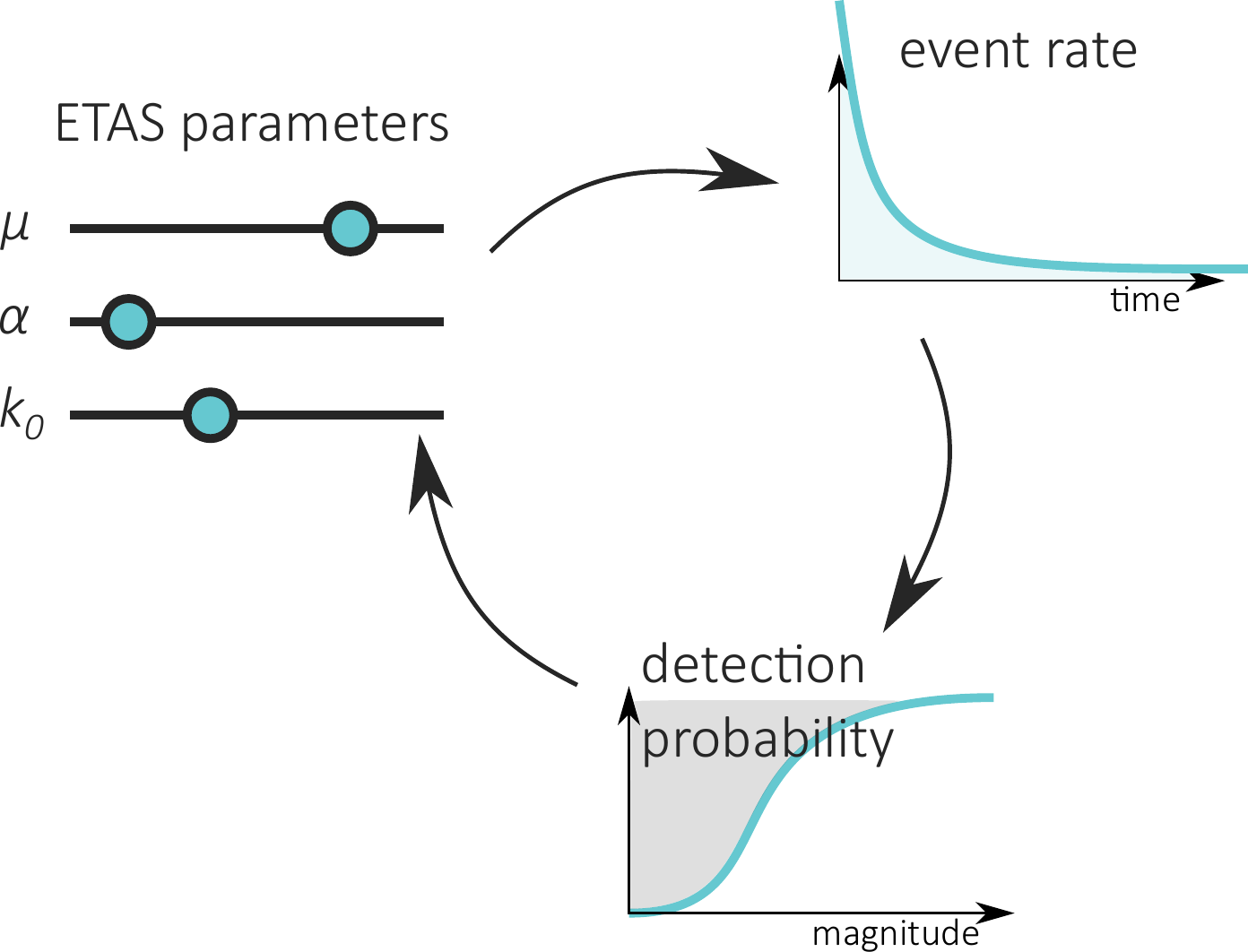}
        \caption{Simplified schematic illustration of PETAI inversion.}
        \label{fig:petai_circular}
    \end{figure}

        The overarching joint inversion of ETAS parameters ($\mathcal{E}$) and high-frequency detection incompleteness ($\mathcal{I} = (\lambda_i, t_R, \beta)$) starts with estimating ETAS parameters in the usual way, i.e. using the algorithm described in Section \ref{sec:zeroone_etas}, with a time-independent completeness magnitude $m_c (=m_{ref})$ above which all events are detected. 
        It then recursively re-estimates $\mathcal{I}$ (see Section \ref{sec:mathcal_I}) and $\mathcal{E}$ (see Section \ref{sec:pdet_etas}) until convergence of the ETAS parameters.
        A simplified illustration of the inversion algorithm is shown in Figure \ref{fig:petai_circular}. Starting with the initial ETAS parameters obtained assuming constant $m_c$, event rates can be calculated at each point in time. Given these event rates, the detection probability function is calibrated, which then provides insight into the temporal evolution of catalog (in-)completeness. ETAS parameters can then be re-estimated, now also using data below $m_c$, by accounting for the estimated incompleteness. With this new set of ETAS parameters, event rates can be re-calculated, upon which detection probability is re-calibrated, and so on, until all convergence criteria are satisfied.
        Figure \ref{fig:flow} shows the detailed flow diagram of the PETAI inversion algorithm.

\section{Synthetic tests}
    \label{sec:synth}
    \subsection{Synthetic Test for ETAS model with long-term variation of $m_c$ (ST1)}
        \label{sec:synth_mc_var}

        To test the ETAS parameter inversion for time-varying $m_c$, we generate 400 complete synthetic catalogs using ETAS and then artificially impose a given $m_c(t)$ on the catalogs. 
        Assuming $m_c(t)$ to be known, we use the method described in Section \ref{sec:zeroone_etas} to infer the parameters used in the simulation.
        
        We estimate $m_c(t)$ based on the Californian catalog described in Section \ref{sec:data} with a time horizon from 1932 to 2019. Fixing the $b$-value we had estimated for the main catalog (1970 - 2019, $M\ge 3.1$, $b = 1.01\pm 0.006$, see \textcite{mizrahi2021effect} for the method used), we estimate $m_c$ for successive 10 year periods starting in 1932. The last period then comprises only 8 years of data.
        Estimation of $m_c$ is analogous to the main catalog, using the method of \textcite{mizrahi2021effect} with an acceptance threshold of $p = 0.1$, but keeping $b = 1.01$ fixed.
        
        This yields
        
        \begin{equation}
            \label{eq:mc_t}
            m_c(t) = \begin{cases}
                4.3 & \text{for $t$ between 1932 and 1941},\\
                3.9 & \text{for $t$ between 1942 and 1951},\\
                4.3 & \text{for $t$ between 1952 and 1961},\\
                3.4 & \text{for $t$ between 1962 and 1971},\\
                3.1 & \text{for $t$ between 1972 and 1981},\\
                3.3 & \text{for $t$ between 1982 and 1991},\\
                2.4 & \text{for $t$ between 1992 and 2001},\\
                2.8 & \text{for $t$ between 2002 and 2011},\\
                3.6 & \text{for $t$ between 2012 and 2019}.
            \end{cases}
        \end{equation}
        
        The large increase in $m_c$ for the years 2012 to 2019 is due to the Ridgecrest events in 2019.
        Although the period affected by aftershock incompleteness only makes up a small fraction of the 8 year period, our method with an acceptance threshold of $p=0.1$ yields a conservative estimate of $m_c$.
        To avoid such an effect, one could use shorter than 10 year periods, or use different methods to estimate time-varying $m_c$.
        
        Note that our method to invert ETAS parameters for time-varying $m_c$ (Section \ref{sec:zeroone_etas}) accepts $m_c(t)$ as an input and works independently of how this $m_c(t)$ was obtained.
        We here want to keep the focus on the parameter inversion and thus choose the described approach to estimate $m_c(t)$ due to its simplicity.
        
        To mimic a realistic scenario, we simulate the synthetic catalogs using parameters obtained after applying ETAS parameter inversion for time-varying $m_c$ on the California data, with two manual corrections.
        
        The first correction is done because it has been shown that certain assumptions in the ETAS model such as a spatially isotropic aftershock distribution or a temporally stationary background rate, as well as data incompleteness can lead to biased estimations of the productivity exponent (\cite{hainzl2008impact}; \cite{hainzl2013impact}; \cite{seif2017estimating}). 
        This bias can lead to a lack of clustering when catalogs are simulated. 
        We thus use an artificially increased productivity exponent $\alpha'$ for our simulations as follows.
        
        Consider the branching ratio $\eta$, defined as the expected number of direct aftershocks (larger than $m_{ref}$) of any earthquake larger than $m_{ref}$,
        
        \begin{equation}
                \eta = \int_{m_{ref}}^{\infty} f_{GR}(m) \cdot G(m) \,dm.
        \end{equation}
        It follows easily that
        
        \begin{equation}
        \label{eq:eta}
            \eta = \frac{\beta \cdot k_0 \cdot \pi \cdot d^{-\rho} \cdot \tau^{-\omega} \cdot e^{c / \tau} \cdot \Gamma(-\omega, \sfrac{c}{\tau})}
        { \rho \cdot (\beta - (a - \rho\gamma))},
        \end{equation} 
        if $\beta>a-\rho\cdot\gamma$, where $\Gamma(s, x)=\int_x^{\infty}t^{s-1}e^{-t}\,dt$ is the upper incomplete gamma function.

        We fix $\alpha'=2.0$ (based on \textcite{helmstetter2003earthquake}; \textcite{guo1997statistical}) and from this derive new values for $a$ and $k_0$, keeping the branching ratio $\eta$ constant. In particular, we define
        
        \begin{align}
            a' &\defeq \alpha' + \rho\cdot\gamma, \label{eq:corr_a}\\
            {k_0}' &\defeq k_0 \cdot \frac{\beta-( a'-\rho\cdot\gamma)}{\beta-(a-\rho\cdot\gamma)} \label{eq:corr_k}.
        \end{align}
        It can be easily shown that in this way, the branching ratio $\eta$ remains the same as long as $\beta-(a-\rho\cdot\gamma)>0$.
        
        Secondly, we reduce the background rate $\mu$.
        In this way, the size of the simulated catalogs is reduced such that inversion requires a reasonable amount of computational power, even for large regions and time horizons.
        The final parameters used for the simulation of the catalogs can be found in Figure \ref{fig:hist_uncert} (black crosses).
        
        400 catalogs of events of magnitude $M\ge 2.4 = m_{ref}$ are simulated as described in Text S1 for the time period of January 1832 to December 2019 in a square of 40$\degree$ lat $\times$ 40$\degree$ long.
        Because of missing long-term aftershocks in the beginning of the simulated catalogs, we allocate a burn period of 100 years in the beginning of the simulated period and are left with catalogs from 1932 to 2019.
        The starting year of our synthetic catalogs coincides with the introduction of instrumentation in California (\cite{felzer2007appendix}).
        This allows us to impose the $m_c(t)$ history observed in California on the synthetic catalogs by discarding all events $e_i$ for which $m_i < m_c(t_i)$.

        We apply the ETAS inversion for time-varying $m_c$ with the here-obtained $m_c(t)$ (see Equation \ref{eq:mc_t}) to the synthetic catalogs. 
        
    \subsection{Synthetic test for PETAI (ST2)}
        \label{sec:meth_synth_test}
        To test the PETAI inversion algorithm, 500 synthetic catalogs are created as follows. 
        We use the parameters obtained after applying the PETAI inversion algorithm to the California data (1970 to 2019) with $m_{ref}$ = 2.5. The value of $m_{ref}$ is chosen to achieve a balance between the amount of data available for the inversion and the computational power required to process such an amount of data.
        For the reasons described in Section \ref{sec:synth_mc_var}, we reduce the background rate $\mu$ and modify the parameters to obtain a corrected productivity exponent as described in Equations \ref{eq:corr_a} - \ref{eq:corr_k}. 
        The final parameters used for the simulation of the catalogs can be found in Figure \ref{fig:synth_results}(d) - (l).
        
        Using these parameters, we simulate as described in Text S1, 500 synthetic catalogs that resemble the Californian catalog, for the period between 1850 and 2020 in a square of 40$\degree$ lat $\times$ 40$\degree$ long.
        As in the previous case, because of missing long-term aftershocks in the beginning of the simulated catalogs, we discard the first 100 years of data and are left with catalogs from 1950 to 2020.
        For each of these catalogs and given the ETAS parameters used for simulation, we calculate the current event rate at the time of each event in the catalogs (Equation \ref{eq:currentrate}).
        As the current event rate is to a large extent driven by aftershock rates of earlier events, we expect overestimation of detection probabilities, as well as overestimation of independence probabilities, during the beginning of the time period (\cite{wang2010missing}; \cite{schoenberg2010relationship}; \cite{nandan2019magnitude}).
        For this reason, we allocate another 20 years of burn period, leaving us with catalogs starting in 1970.
        
        Each of the 500 catalogs are then artificially made incomplete as follows. Using the detection probability function given by Equation \ref{eq:defprob}, and the $b$-value of 1.03 estimated from the Californian catalog using PETAI inversion, we calculate for each event its probability of being detected. According to this probability we randomly decide for each event whether it has been detected or not. The subset of all events that were detected is then used as a test catalog. This is done assuming different values for $t_R$ of 1.97 (as estimated from the Californian catalog), 5, 10, 30, 60, and 180 minutes, yielding six variations of the test catalog per originally simulated catalog, which makes a total of 3000 test catalogs. 
        The value of $t_R$ greatly influences the fraction of undetected events in the resulting catalog, and we chose to investigate different values of $t_R$ to ensure there are sufficiently many test catalogs with a fraction of undetected events similar to the fraction of estimated undetected events inferred for California.
        This estimated number of undetected events is obtained by summing $\zeta(t_i)$, the expected number of unobserved events per observed event, which is estimated as a component of the PETAI inversion, over all occurrence times $t_i$ of events in the primary catalog.
        
        \subsection{Results for ST1}
        \label{sec:Results_for_ST1}
        Figure \ref{fig:hist_uncert} shows the ETAS parameters used in the simulation of the synthetic catalogs, and the median, distribution, and 95\% confidence intervals of the parameters inverted from the synthetic catalogs.
        The parameters estimated from the synthetic catalogs lie reasonably close to the data-generating parameters.
        In particular, $a$, $c$, $\omega$, $\tau$ and $\gamma$ are accurately inverted, while $\mu$, $k_0$, $d$ and $\rho$ tend to be overestimated.
        The reason for the overestimation of $\rho$ stems from a computational simplification made during inversion. 
        In order to avoid extremely large triggering probability matrices, we only consider pairs of source and target events with a spatial distance of less than 50 source lengths, where one source length is defined using the magnitude to length scaling relations defined in \textcite{wells1994new}. 
        This upper limit for distances between event pairs translates to an exaggeration of the estimated values of $\rho$. 
        We confirmed that as we gradually relax the cutoff criterion, the estimated value of $\rho$ moves closer to the true values used for generating the synthetic catalog. 
        The regularizer of the spatial kernel, $d$, is positively correlated with $\rho$, hence an overestimation of the latter translates to an overestimation of the former. 
        The overestimation of $\mu$ can also be explained, considering that distant aftershocks have a higher tendency to appear independent due to the artificially imposed cutoff criterion.   
      
        \subsection{Results for ST2}
            \subsubsection{Inverted number of undetected events}
                
                Figure \ref{fig:synth_results} (a) shows the series of events of one example synthetic test catalog over the primary time period in blue, with the undetected synthetic events marked in black. 
                The number of undetected events is 1282, which makes up 6.25\% of the original synthetic catalog.
                Figure \ref{fig:synth_results} (b) shows cumulative number of undetected synthetic events over time in black, compared to the cumulative inferred number of undetected events in blue for the same example catalog.
                Overall, it is estimated as a result of applying the PETAI inversion that 1068.88 events were undetected in the example catalog. While this underestimates the true number of 1282 undetected events, the major part of events can be reconstructed, with accurate timing.
                
                Figure \ref{fig:synth_results} (c) shows inferred versus actual number of undetected events for 3000 test catalogs assuming different detection efficiencies.
                The estimated fraction of undetected events is distributed around the actual fraction of undetected events, and the median estimated fraction matches well the median actual fraction, with a slight tendency towards underestimation.
                
            \subsubsection{Accuracy of inverted parameters}
                \label{sec:res_syn_petai_par}
                Figure \ref{fig:synth_results} (d) - (n) shows the ETAS parameters and $(t_R, \beta)$ that were used in the simulation of the synthetic catalogs, and the parameters inverted from these synthetic catalogs.
                In general, the inverted parameters correspond well to the parameters used in the simulation, although some of the estimates are slightly biased.
                The parameters $c$ and $\omega$, both describing the temporal decay of aftershock rate, show a trend of increasing bias with increasing $t_R$, that is, with increasing incompleteness. 
                For the other parameters, no clear dependency of the bias on $t_R$ is recognizable.
                The estimate of $c$ matches the true value almost perfectly for $t_R$ = 1.97 minutes, but starts being overestimated for larger values of $t_R$ above 30 minutes.
                On the other hand, $\omega$ shows an increasing tendency of being underestimated with increasing values of $t_R$. 
                Earlier aftershocks have a larger tendency to be missing due to STAI, which leads to a seemingly slower decay of aftershock rate in time. 
                As the PETAI algorithm has a tendency to underestimate STAI (Figure \ref{fig:synth_results} (c)), and this tendency increases with increasing $t_R$, this translates into an increasing negative bias in the inferred values of $\omega$.
                
                Qualitatively, the tendencies to over- or underestimate the remaining parameters are identical with the tendencies observed in ST1 (Section \ref{sec:Results_for_ST1}).
                It is therefore plausible that these tendencies are consequences of a finite time horizon and finite spatial window used in the simulation of the synthetics, rather than being artifacts of the PETAI inversion algorithm.
                
                
        
                Finally, we observe a tendency to underestimate $t_R$, which means that detection probabilities tend to be overestimated. This is in line with our previous observation that the fraction of undetected events tends to be slightly underestimated, suggesting the PETAI inversion to be slightly conservative.

\section{Application to California}
    \label{sec:applied}
        We calculate ETAS parameters, $\beta$ and $t_R$ (if applicable)  using different inversion algorithms to Californian data.
        Additionally, we provide the resulting values for productivity exponent $\alpha = a - \rho \gamma$ and branching ratio $\eta$ (see Equation \ref{eq:eta}).
        
        First, we apply usual inversion method as described in Section \ref{sec:zeroone_etas} with a constant completeness magnitude of $m_c\equiv 3.1$ to the main catalog (1970 to 2019).
        Then, we invert the parameters by accounting for long-term time-variation of completeness (Equation \ref{eq:mc_t}).
        In this case, the extended catalog from 1932 to 2019 can be used with a reference magnitude of $m_{ref} = 2.4$.
        Finally, we apply PETAI inversion to the main catalog (1970 to 2019) with a reference magnitude of $m_{ref} = 2.5$.
        Note that the estimation of $\beta$ is independent of the ETAS parameter estimates for the first two applications, but not in the case of PETAI inversion (see Section \ref{sec:petai}).
        
        To allow a better comparison between parameters inverted using different methods when $m_{ref}$ varies, we translate the parameters to a reference magnitude of $m_{ref} = 3.1$ as follows.
        With the exception of $\mu, k_0$ and $d$, all parameters are $m_{ref}$-agnostic, and the three exceptions can easily be adjusted. Denote by $\Delta m$ the difference between new and original reference magnitude, $\Delta m = m_{ref}'- m_{ref}$. Then,
        
        \begin{equation}
            \label{eq:transform_d}
            d' \defeq d \cdot e^{\Delta m \cdot \gamma}
        \end{equation}
        ensures that
        
        \begin{equation}
            d \cdot e^{\gamma\cdot(m - m_{ref})} = d' \cdot e^{\gamma\cdot(m - m_{ref}')}.
        \end{equation}
        Stipulating that the branching ratio $\eta$ (Equation \ref{eq:eta}) remains unchanged, it follows that
        
        \begin{equation}
            k_0' \coloneqq k_0 \cdot e^{\Delta m \cdot \gamma \cdot \rho}.
        \end{equation}
        The adaptation of the background rate $\mu$ follows trivially from the GR law,
        
        \begin{equation}
            \label{eq:transform_mu}
            \mu' = \mu \cdot e^{-\beta \cdot \Delta m}.
        \end{equation}
        \\
    
        \subsection{Interpretation of inverted parameters}
        \label{sec:parameter_comm}

        Table \ref{tab:params_cali} shows the estimated values of ETAS parameters, $\beta$, and $t_R$ (if applicable) obtained using different inversion algorithms to Californian data.
        Additionally, the resulting values for the productivity exponent $\alpha = a - \rho \gamma$ and branching ratio $\eta$ (see Equation \ref{eq:eta}) are provided.
        The first, second, and fourth column show the parameters obtained from applying the method with $m_c\equiv 3.1$, when using long-term-variations of $m_c$, and when using PETAI, respectively.
        Columns three and five contain the parameters of columns two and four after having been transformed to a reference magnitude of $m_{ref} = 3.1$.
        
        Overall, the inverted parameters are roughly consistent among the three algorithms. 
        Although there are slight differences between the estimated parameters, they can plausibly be attributed to different input datasets, which vary for the three algorithms in either time-span or magnitude range. 
        In the following, we present some speculative explanations of the observed differences.
            
        We find that the estimate of $\tau$ obtained from the ETAS model calibrated on the extended catalog (1932 to 2019) with the long-term time variation of $m_c$ is smaller than in the other two cases, to an extent that the uncertainties obtained in the synthetic tests cannot explain this decrease. 
        This decrease despite the use of a catalog spanning a longer duration compared to the other two cases, shows that $\tau$ may actually better reflect the long-term behavior of earthquake interaction, rather than being determined by the finite duration of the catalog. 
        Note that if the temporal finiteness of the catalog was the dominant factor in the determination of $\tau$, one would expect an increase of $\tau$ with increasing time spanned by the catalog.
        Furthermore, the less pronounced decrease of $\tau$ in case of the PETAI inversion speaks against the possibility that the decrease is caused by inclusion of lower magnitude earthquakes revealing previously unseen earthquake interactions.
        
        A somewhat counter-intuitive observation is the increase of $c$ for both new inversion techniques. For the case of long-term variation of $m_c$, in particular, $c$ shows a significant increase considering the expected uncertainties.
        The parameter $c$ has been interpreted to reflect aftershock incompleteness (\cite{kagan2004short}; \cite{lolli2006comparing}; \cite{hainzl2016apparent}) and would thus be expected to decrease when this effect is accounted for by the model (\cite{seif2017estimating}).
        The observed higher value of $c$ even after accounting for STAI thus requires a different interpretation of $c$.
        \textcite{narteau2009common} found a dependency of $c$ on faulting style, and brought the parameter in relation with differential stress and the intensity of stress re-distribution.
        Another possible interpretation provided by \textcite{lippiello2007dynamical} is based on the dynamical scaling hypothesis in which time differences relate to magnitude differences.
        \textcite{shcherbakov2004generalized} proposed a generalized Omori law which incorporates three empirical scaling laws (\cite{gutenberg1944frequency}, \cite{baath1965lateral}, \cite{utsu1961statistical}) with a dependence of $c$ on the cutoff magnitude which can qualitatively explain our observations: The value inverted for $c$ is highest in the case of $m_{ref} = 2.4$, and lowest for $m_{ref} = 3.1$.
        Overall, one should be careful to not over-interpret this estimate of $c$.
        After all, $c$ is overestimated for large values of $t_R$ in the PETAI synthetic test and hence an observed increase in $c$ might be a consequence of complex interdependencies of all parameters involved.
        
        While the branching ratio $\eta$ does not substantially vary with the different inversion methods, we observe a slightly increased productivity exponent for the PETAI inversion.
        Although the increase lies within expected uncertainty, such an increase is expected given the results of \textcite{seif2017estimating}, with the extent of the observed increase being in line with their estimated extent of underestimation for the productivity exponent.
        
        The background rate $\mu$ shows a significant increase when a longer time horizon is considered, and decreases significantly when STAI is accounted for.
        As $\mu$ is clearly overestimated in the synthetic test with long-term variation of $m_c$, and only slightly overestimated in the case of PETAI, we may suspect that the increased value for $\mu$ in the first case is an artifact of the inversion method, while the decrease in background rate with PETAI could suggest that including smaller magnitude events in our model by accounting for incompleteness reveals previously hidden earthquake interactions, resulting in a lower $\mu$.
        
        The parameter $\gamma$, which describes the exponential relationship between earthquake magnitude and the distance to the event at which the aftershock rate starts to decrease faster, is significantly increased in the case of long-term variation of $m_c$.
        Slight overestimation is expected based on the synthetic tests, but not to this extent.
        
        At the same time, $\rho$ increases for both new inversion techniques.
        Again, overestimation of $\rho$ is expected given the results of the synthetic tests and the observed values might thus be artifacts of the algorithms applied.
        As the problem of the finite spatial region applies in the same way to standard ETAS as well as the other two methods, this is unlikely to be the cause of the difference in parameter estimates.
        
        The value of $\beta$ shows an increase from 2.33 to 2.37, which translates to a $b$-value increase from 1.01 to 1.03, when STAI is accounted for in the PETAI inversion. 
        This is expected due to the underestimated number of small events caused by STAI.
 
    \subsection{Incompleteness insights through PETAI}
        In addition to a new set of estimated ETAS parameters, applying the PETAI inversion to the Californian catalog produces further interesting outputs.
        Similarly to the case of the synthetic catalog, Figure \ref{fig:cali_petai} (a) shows the estimated cumulative number of undetected events over time.
        As expected, the increase is predominantly step-wise, caused by short, incomplete periods during aftershock sequences, and long, complete periods in-between.
        While the total expected number of undetected events is at 5041.74, the extrapolated number obtained from a GR law fitted on $M\ge 3.1$ events is only 88.91.
        This estimate of the number of unobserved events differs from the PETAI estimate in that it assumes perfect detection above $M 3.1$.
        Although the true number of undetected events can never be known, the synthetic test suggests that the PETAI result is reliable and even slightly conservative, and thus the GR law extrapolation would be a severe underestimation of the true number of undetected events.

        The magnitude-dependent detection probability evolution is illustrated in Figure \ref{fig:cali_petai} (b).
        In around 84\% of event times $t_i$, events of magnitude $M\geq 4$ are expected to be detected with a probability of 99.9\% or more.
        Similarly, in 82\% of event times $t_i$, $M\geq 3$ events are expected to be detected with a probability of 99\% or more.
        Spikes of incompleteness during large sequences lead to detection probabilities of less than 50\% for smaller events, in the most extreme case for events of magnitude $M\leq 3.47$.
        
        As expected, periods of elevated incompleteness coincide with the periods of rapid increase in undetected events shown in (a).
        The last step in (a), which corresponds to the 2019 Ridgecrest sequence, is extraordinarily large  compared to all previous steps.
        This is most likely explained by the fact that the sequence was better recorded than comparable sequences in previous years.
        When the detection capability of the seismic network improves, the recovery time $t_R$ becomes shorter.
        Because we have assumed $t_R$ to be stationary for simplicity, a larger number of recorded events will lead to a smaller estimated detection probability, which in turn leads to larger numbers of expected undetected events.
        In future versions of the model, to avoid such artifacts, it would be advisable to combine the possibility of including long-term changes in completeness (as in the model described in Section \ref{sec:zeroone_etas}) with rate-dependent aftershock incompleteness by means of a non-stationary $t_R$.
              
        Figure \ref{fig:cali_petai} (c) - (g) shows excerpts of Figure \ref{fig:cali_petai} (b) for the 1989 M6.9 Loma Prieta, the 1992 M6.1 Joshua Tree and 7.3 Landers, the 1994 M6.7 Northridge, the 1999 M7.1 Hector Mine, and the 2019 M6.4 and 7.1 Ridgecrest events, in comparison to the $m_c(t)$ estimate given by the formulation of \textcite{helmstetter2006comparison} which was provided for Southern California.
        While their definition is not probabilistic, we observe that their $m_c$ 5 minutes after the mainshock lies between 90\% and 99\% detection according to PETAI.
        The shape of the recovery from incompleteness does not fully coincide for the two methods, with generally slower recovery in the case of PETAI for the shown excerpts.
        \textcite{helmstetter2006comparison} use a simpler formulation, and do not provide arguments for their specific choice of parameterization of $m_c(t)$.
        On the other hand, the parametric description of the magnitude of 50\% detection by \textcite{ogata2006immediate}, presented for the example of the 2003 Miyagi-Ken-Oki earthquake, takes a shape similar to the one obtained through PETAI, although it is not quantitatively comparable to the case of California.
        
        The range of observed states of detection efficiency during the 2019 Ridgecrest sequence is visualized in Figure \ref{fig:cali_petai} (h).
        Prior to the large events, detection is almost perfect for all magnitudes.
        After the M6.4 event, detection is weakened and recovers with time, until the M7.1 mainshock, when it is again weakened.
        Around 15 minutes after the earthquake, events of magnitude below 3.0 still have almost no chance to be detected, with M3.5 events having roughly a 50\% chance to be detected.
        After three hours, detection has already clearly improved, although M2.5 events are still almost surely not detected.
        After six days, the detection probability function almost corresponds to the prefect detection state, which was in place prior to the main events.
        

    \subsection{Comments on computational time}
        \label{sec_comp_time}
            There are two aspects to consider when discussing the computational time of the parameter inversion techniques presented here.
            On one hand, the increased complexity of the algorithms plays an important role.
            In particular, the PETAI inversion comprises multiple loops of ETAS and incompleteness estimation.
            Although convergence was usually reached after 4 iterations, this still implies a minimum factor of 4 in terms of computation time which is only required for ETAS inversion, on top of which comes the time needed for the estimation of detection parameters and event rates.
            The second factor, which contributes even more to an increase of computation time, is the increased size of the catalog which is available to be used.
            For our application to Californian data, the number of events used in the PETAI inversion increases by a factor of 3.78 because the minimum considered magnitude is reduced from 3.1 to 2.5.
            The leads the number of pairs of potentially related events to increase from 7.3 million to 47.1 million. Note that these numbers are obtained after imposing the 50 source length cutoff criterion described in \ref{sec:Results_for_ST1}.
            While this increase in the number of potentially related event pairs causes a substantial increase in run time, educated initial guesses for ETAS parameter inversions can substantially reduce run time without affecting the results.
            Our Python 3.8 implementation of the PETAI inversion, run with a single core (Intel Xeon E5-2697v2) of the Euler high-performance computing (HPC) cluster at ETH Zurich, took 23 hours. 
            Roughly 20\% of this time was spent on the optimization of event rates and detection parameters, and 80\% on the optimization of ETAS parameters.
            
            In contrast to the PETAI inversion, the run time of the ETAS parameter inversion with time-varying $m_c$ is barely affected by model complexity.
            During synthetic experiments, we found the run time to be comparable to the run time of the usual ETAS inversion when the number pairs of potentially related events was similar.
    
\section{Pseudo-prospective forecasting experiments}
    \label{sec:pseudo}
        To better understand if and how the PETAI model can improve earthquake forecasts, we conduct pseudo-prospective forecasting experiments.
        Note that as these experiments are computationally expensive, we conduct them for the PETAI method only.
        As most aftershocks occur soon after their triggering event, accounting for STAI in ETAS simulations seems promising for forecasting.
        The parameter inversion for long-term variations of $m_c$ is mainly intended as a tool to obtain ETAS parameters in regions where data is sparse and a model inversion would not be possible otherwise.

        \subsection{Competing models}
            We compare five models.
            \begin{enumerate}
                \item The base ETAS model assumes perfect detection above a constant $m_c=3.1$ and is used as the null model.
                
                \item PETAI, the alternative model, has two modifications to the null model.
                First, it uses improved ETAS parameter estimates that were obtained in the PETAI inversion with a reference magnitude $m_{ref}$ of 2.5.
                Second, magnitude $M\ge 2.5$ earthquakes are allowed to trigger and be triggered. 
                For this, the events in the training catalog, which act as triggering earthquakes in the simulation, have their triggering capability inflated by $1 + \xi(t)$, as estimated in the PETAI inversion.
            \end{enumerate}
            Two intermediate models are assessed to dissect the effect of the two modifications.
            \begin{enumerate}
                \setcounter{enumi}{2}
                \item par\_only uses ETAS parameter estimates obtained from PETAI, but only $M\ge 3.1$ events are allowed to trigger and be triggered, assuming perfect detection there (i.e. $\xi(t)\equiv 0$).
                In this case, the parameters obtained for the PETAI model have to be transformed to be compatible with a reference magnitude of $m_{ref} = 3.1$ as described in Equations \ref{eq:transform_d} - \ref{eq:transform_mu}.
                
                \item Vice-versa, trig\_only allows $M\ge 2.5$ events to trigger and be triggered, using the inverted $\xi(t)$ for inflated triggering, but does not use the improved ETAS parameter estimates.
                In this case, the parameters obtained for the null model have to be transformed to be compatible with a reference magnitude of $m_{ref} = 2.5$ as described in Equations \ref{eq:transform_d} - \ref{eq:transform_mu}.
            \end{enumerate}
            Lastly, we assess an additional benchmark model to test whether deliberately underestimating $m_c$ is an appropriate alternative to the rather complex PETAI model.
            \begin{enumerate}
                \setcounter{enumi}{4}
                \item low\_mc assumes perfect detection above a constant $m_c=2.5$. This model uses neither the parameter estimates obtained from PETAI, nor the inverted $\xi(t)$ for inflated triggering, but it allows $M\ge 2.5$ events to trigger and be triggered and thus is based on the same data as the PETAI-based models.
            \end{enumerate}

        \subsection{Experiment setup}
            For a testing period length of 30 days, we define a family of training and testing periods such that the testing periods are consecutive and non-overlapping. Each training period ends with the starting date of its corresponding testing period. The starting date of the first testing period is January $1^{st}$, 2000. The end date of the last of the 244 testing periods is January $16^{th}$, 2020.
            
            For each testing period, all competing models are trained based on the corresponding training data.
            Then, forecasts are issued with each model through simulation of 100,000 possible continuations of the training catalog. 
            Because the testing data is ignored when the models are calibrated, these forecasts are pseudo-prospective.
            This is done by simulating Type I earthquakes (the cascade of aftershocks of earthquakes in the training catalog) and Type II earthquakes (simulated background earthquakes and their cascade of aftershocks) similarly to how it is described by \textcite{nandan2019forecastingfull}.
            The algorithm for simulation is described in detail in Text S1.
            
            The performance of each model is evaluated by calculating the log-likelihood of the testing data given the forecast. See Text S2 for details on the calculation of the log-likelihood using the full distribution approach as described by \textcite{nandan2019forecastingfull} for a fair evaluation of ETAS-based models. 
            Two competing models can be compared by calculating the information gain (IG) of the alternative model $M_{alt}$ over the null model $M_0$, which is simply the difference in log-likelihood of observing the testing data.
            The mean information gain (MIG) is calculated as the mean over all testing periods.
            This evaluation metric is similar to other metrics that have been used for model comparison, such as the total information gain or information gain per earthquake (IGPE) used in the CSEP T-test (\cite{harte2005entropy}; \cite{rhoades2011efficient}; \cite{zechar2013regional}; \cite{strader2017prospective}, see \cite{savran2020pseudoprospective} for recent complementary CSEP testing metrics) or the residual-based log-likelihood ratio score (\cite{clements2011residual}; \cite{bray2014voronoi}; \cite{gordon2015voronoi}; \cite{gordon2021nonparametric}).
            
            As an additional benchmark, we calculate the total IGPE of the ETAS null model versus a spatially and temporally homogeneous Poisson process (STHPP) model.
            Note that the STHPP model is not considered a participant of the forecasting experiment and superiority is always discussed relative to the ETAS null model.
            
            For details on the STHPP model and on the conditions under which one model is considered superior over another, see Text S2 and \textcite{nandan2019forecastingfull}.

        \subsection{Time evolution of the parameters of the competing models}
        Figure \ref{fig:par_evol_fc} shows the parameter evolution with increasing training period obtained with standard ETAS ($m_c=2.5$ and $m_c=3.1$) and PETAI inversion.
        Two parameters, namely $\mu$ and $\tau$, show a systematic decrease and increase, respectively, with growing time horizon of the training catalog.
        When compared to the uncertainties in the synthetic tests, the extents of the changes of $\mu$ are larger than the 95\% confidence intervals, while the changes of $\tau$ lie within the expected uncertainties.
        A possible explanation for this observation is that an increased time horizon of the training catalog reveals more long-term earthquake interactions, leading to a higher value of $\tau$, that is a later onset of the exponential taper in the temporal aftershock density, and simultaneously to a lower background rate $\mu$, as more events can be interpreted as aftershocks of previous earthquakes.
        
        Nearly all parameter estimates show a jump in 2010, caused by the 2010  El-Mayor Cucapah earthquake sequence, and a second jump in 2019, caused by the 2019 Ridgecrest sequence.
        There are several reasons why such jumps in parameter estimates could occur.
        In the case of the 2010 events, the main earthquake occurred outside of California and thus network coverage can play a role, as well as the absence of a large fraction of aftershocks due to the boundaries of the considered region.
        Furthermore, triggering parameters can differ between regions, sequences and can also depend on the magnitude of the mainshocks (\cite{nandan2019magnitude}; \cite{nandan2021global}; \cite{ouillon2005magnitude}; \cite{sornette2005multifractal}; \cite{nandan2021triggering}). These dependencies can increase the representation of the active region and particular sequences in the catalog and lead to sudden changes in the overall parameters.
        
        \subsection{Forecasting performance of the competing models}
        Figure \ref{fig:fc_results} shows the results of the pseudo-prospective forecasting experiments.
        For a target magnitude threshold of $m_t = 3.1$, PETAI as well as trig\_only significantly outperform the ETAS null model with p-values of virtually 0 and a mean information gain of 0.97 and 0.94, respectively.
        Note that this improvement is over a very strong null model, which has a total information gain of 49'246 (i.e. a  MIG of 202.66 or an IGPE of 5.62) over the STHPP model.
        PETAI has a slightly positive but not statistically significant information gain compared to trig\_only.
        On the other hand, par\_only and low\_mc do not significantly outperform the ETAS null model.
        This suggests that the main driver of the improvement of the forecast is the inclusion of small events between $M 2.5$ and $M 3.1$ in the simulations, rather than the newly obtained parameter estimates.
        It also indicates that accounting for incompleteness, which is possible due to the estimated $\xi(t_i)$ obtained in the PETAI inversion, is necessary for this improved forecast.
        The sole inclusion of events between $M 2.5$ and $M 3.1$ in the simulations assuming completeness above $M 2.5$ is not sufficient to obtain significant improvements.
        For all considered values of $m_t$, PETAI and trig\_only rank higher than low\_mc in terms of MIG, which further supports the idea that accounting for STAI is relevant for improved ETAS-based earthquake forecasting.
        
        The temporal evolution of the cumulative information gain of the two superior models shows a decrease during the 2010 El Mayor-Cucapah and the 2019 Ridgecrest sequences.
        Those sequences were most active in Southern California, where the seismic network is much denser than in the rest of the considered region (\cite{hutton2010earthquake}; \cite{schorlemmer2008probability}).
        The assumption of spatially homogeneous detection incompleteness is thus inaccurate and may be the reason for over-inflation of the aftershock productivity during these sequences, explaining the decrease in information gain.
        One can therefore expect that accounting for spatial variation of STAI in subsequent models may lead to even better forecasts. 
        
        With increasing values of $m_t$ to 3.5, 4.0, 4.5, and 5.0, the IGPE of the ETAS null model versus the STHPP model increases to 5.92, 5.92, 6.62, and 7.44, respectively.
        At the same time, the mean information gain values between the competing ETAS-based models generally decrease, and almost no model significantly outperforms any other competing model.
        Occasionally, par\_only is outperformed by the ETAS null model or by trig\_only.
        These observations suggest that taking into account information about smaller earthquakes mainly helps improving ETAS-based forecasts of smaller earthquakes.
        More precisely, simulating aftershocks of small earthquakes is the key ingredient for improved forecasting of similarly-sized events.
        Although within the framework of the standard ETAS model, small earthquakes can trigger large ones, and their relative abundance implies significant contribution to the overall triggering (\cite{marsan2005role}; \cite{helmstetter2005importance}; \cite{sornette2005constraints}), we find that the beneficial effect vanishes when forecasting large events. Additional ways exist in which small earthquakes can contribute to improving forecasting models. Besides their potential to cumulatively contribute to aftershock triggering, the large number of earthquakes below $m_c$ can help to highlight the underlying fault structure, which, when accounted for, can significantly improve forecasting performance (\cite{gordon2021nonparametric}; \cite{bach2012improving}; \cite{cattania2018forecasting}; \cite{guo2015improved}). In fact, small earthquakes have been shown to improve forecasts in the context of other models (\cite{mancini2019improving}; \cite{mancini2020predictive}), and somewhat mixed overall results but a clear signal that small earthquakes do contribute to triggering through the redistribution of static stresses have been reported (\cite{meier2014search}; \cite{segou2013comparative}; \cite{nandan2016systematic}).
        
        \textcite{helmstetter2006comparison} compared the probability gain of their time-dependent model versus their similar but time-independent model and found that probability gain decreases with an increasing target magnitude threshold.
        They speculated that this observation may be due to a smaller sample size when the target magnitude threshold increases.
        \textcite{helmstetter2014adaptive} found the same decrease in the context of a different, non-parametric kernel space-time smoothing model.
        Although the study was based on a larger amount of data than \textcite{helmstetter2006comparison}, they likewise speculated that this decrease is due to a small sample size. 
        In our case, the same effect is observed at considerably large sample sizes of 3601, 1111, 307, and 85 events for $m_t = 3.5, 4.0, 4.5,$ and $5.0$.
        Another possible explanation for this effect is provided by the findings of multiple previous studies using both non-parametric (\cite{nichols2014assessing}; \cite{spassiani2016exploring}) and parametric (\cite{nandan2019magnitude}; \cite{nandan2021triggering}) approaches, that earthquakes tend to preferentially trigger aftershocks of similar size.
        Their results can explain the improved forecast of small events when small events are used for simulation, as well as the vanishing of this improvement when the magnitude difference between newly included events and target events becomes large.
        This could furthermore serve as an alternative explanation of the results of \textcite{helmstetter2006comparison} and \textcite{helmstetter2014adaptive}.
        
        Note that the IGPE of the ETAS null model against the STHPP model increases with increasing $m_t$, while \textcite{helmstetter2006comparison} observe a decrease in probability gain against a temporally homogeneous and spatially variable Poisson model. 
        This suggests that spatial inhomogeneity becomes more important with increasing target magnitude, while temporal inhomogeneity based on small earthquakes becomes less important with increasing target magnitude.
        It is important to highlight that the 30-day testing periods of the present study, in contrast to 1-day periods as in \textcite{helmstetter2006comparison}, prevent ETAS models from being updated after large events.
        This likely understates the extent of superiority that could be achieved by models which include $M2.5$ to $M3.1$ events, and thus the role of small earthquakes, in daily forecasts.

\section{Conclusion}
    \label{sec:conclusion}
    
    We propose a modified algorithm for the inversion of ETAS parameters when $m_c$ varies with time, and an algorithm for the joint inversion of ETAS parameters and probabilistic, epidemic-type aftershock incompleteness.
    We test both methods on synthetic catalogs, concluding that they can accurately invert the parameters used for simulation of the synthetics.
    The given formulations are rather general and can equally be applied to spatial or spatio-temporal variations of $m_c$, as well as to any suitable definition of a detection probability function.
    
    Two potential use cases are the estimation of ETAS parameters based on the Californian catalog since 1932 with long-term fluctuations of $m_c$ between 4.3 and 2.4, and the estimation of ETAS parameters and short-term aftershock incompleteness based on the incomplete Californian catalog of events above $M 2.5$.
    The latter is further used to test the forecasting power of small earthquakes. Results of numerous pseudo-prospective forecasting experiments suggest that
    \begin{itemize}
        \item Information about small earthquakes significantly and substantially improves forecasts of similar-sized events.
        \item Main driver of this improvement is the simulation of aftershocks of small events.
        \item Accounting for incompleteness when simulating aftershocks of small events is necessary to achieve this improvement.
        \item Information about small earthquakes does not significantly affect the performance of large event forecasts.
    \end{itemize}
    A possible explanation for these results is provided by previous findings (\cite{nichols2014assessing}; \cite{spassiani2016exploring}; \cite{nandan2019magnitude}; \cite{nandan2021triggering}), that earthquakes preferentially trigger aftershocks of similar size.
    
    Our results have potentially significant implications for the future of earthquake forecasting.
    Thanks to the here-presented algorithms, ETAS models may be calibrated for regions with low seismicity where the usual inversion algorithms would fail due to missing data.
    To facilitate the embracing of data incompleteness in such cases, our inversion codes will be made openly available after publication of the article through github.com/lmizrahi/etas and github.com/lmizrahi/petai.
    
    The newly gained insights from forecasting experiments guide us in the search of the next generation earthquake forecasting models.
    Besides other discussed topics such as anisotropy, temporally or spatially non-stationary background rate (\cite{hainzl2008impact}; \cite{hainzl2013impact}; \cite{nandan2020earth}), the importance of accounting for short-term incompleteness when simulating, as well as a magnitude-dependent distribution of aftershock magnitudes are emphasized.

\acknowledgments
The data used for this analysis is available through the website\\
\noindent https://earthquake.usgs.gov/earthquakes/search/ (\cite{anss}).
The authors wish to thank Sebastian Hainzl, Andrew Michael and Andrea Llenos for insightful discussions and helpful feedback on earlier versions of this article. We also want to thank the editor Rachel Abercrombie, the anonymous associate editor, one anonymous reviewer, and Max Werner for their constructive feedback which greatly improved this article.
This work has received funding from the Eidgenössische Technische Hochschule (ETH) research grant for project number 2018-FE-213, “Enabling dynamic earthquake risk assessment (DynaRisk)” and from the European Union’s Horizon 2020 research and innovation program under Grant Agreement Number 821115, real-time earthquake risk reduction for a resilient Europe (RISE).


%
%

\printbibliography[heading=bibintoc]

@article{ogata1998space,
  title={Space-time point-process models for earthquake occurrences},
  author={Ogata, Yosihiko},
  journal={Annals of the Institute of Statistical Mathematics},
  volume={50},
  number={2},
  pages={379--402},
  year={1998},
  publisher={Springer}
}

@article{nandan2017objective,
  title={Objective estimation of spatially variable parameters of epidemic type aftershock sequence model: Application to California},
  author={Nandan, Shyam and Ouillon, Guy and Wiemer, Stefan and Sornette, Didier},
  journal={Journal of Geophysical Research: Solid Earth},
  volume={122},
  number={7},
  pages={5118--5143},
  year={2017},
  publisher={Wiley Online Library}
}

@article{veen2008estimation,
  title={Estimation of space--time branching process models in seismology using an em--type algorithm},
  author={Veen, Alejandro and Schoenberg, Frederic P},
  journal={Journal of the American Statistical Association},
  volume={103},
  number={482},
  pages={614--624},
  year={2008},
  publisher={Taylor \& Francis}
}

@article{petersen20182018,
  title={2018 one-year seismic hazard forecast for the central and eastern United States from induced and natural earthquakes},
  author={Petersen, Mark D and Mueller, Charles S and Moschetti, Morgan P and Hoover, Susan M and Rukstales, Kenneth S and McNamara, Daniel E and Williams, Robert A and Shumway, Allison M and Powers, Peter M and Earle, Paul S and others},
  journal={Seismological Research Letters},
  volume={89},
  number={3},
  pages={1049--1061},
  year={2018},
  publisher={GeoScienceWorld}
}

@article{wiemer2009probabilistic,
  title={Probabilistic seismic hazard assessment of Switzerland: best estimates and uncertainties},
  author={Wiemer, Stefan and Giardini, Domenico and F{\"a}h, Donat and Deichmann, Nicholas and Sellami, Souad},
  journal={Journal of Seismology},
  volume={13},
  number={4},
  pages={449},
  year={2009},
  publisher={Springer}
}

@article{bach2012improving,
  title={Improving empirical aftershock modeling based on additional source information},
  author={Bach, Christoph and Hainzl, Sebastian},
  journal={Journal of Geophysical Research: Solid Earth},
  volume={117},
  number={B4},
  year={2012},
  publisher={Wiley Online Library}
}

@article{gutenberg1944frequency,
  title={Frequency of earthquakes in California},
  author={Gutenberg, Beno and Richter, Charles F},
  journal={Bulletin of the Seismological Society of America},
  volume={34},
  number={4},
  pages={185--188},
  year={1944},
  publisher={The Seismological Society of America}
}

@article{marzocchi2014some,
  title={Some thoughts on declustering in probabilistic seismic-hazard analysis},
  author={Marzocchi, W and Taroni, M},
  journal={Bulletin of the Seismological Society of America},
  volume={104},
  number={4},
  pages={1838--1845},
  year={2014},
  publisher={Seismological Society of America}
}

@article{nandan2019forecastingrates,
  title={Forecasting the rates of future aftershocks of all generations is essential to develop better earthquake forecast models},
  author={Nandan, Shyam and Ouillon, Guy and Sornette, Didier and Wiemer, Stefan},
  journal={Journal of Geophysical Research: Solid Earth},
  volume={124},
  number={8},
  pages={8404--8425},
  year={2019},
  publisher={Wiley Online Library}
}

@article{nandan2019forecastingfull,
  title={Forecasting the Full Distribution of Earthquake Numbers Is Fair, Robust, and Better},
  author={Nandan, Shyam and Ouillon, Guy and Sornette, Didier and Wiemer, Stefan},
  journal={Seismological Research Letters},
  volume={90},
  number={4},
  pages={1650--1659},
  year={2019},
  publisher={Seismological Society of America}
}

@article{nandan2019magnitude,
  title={Magnitude of Earthquakes Controls the Size Distribution of Their Triggered Events},
  author={Nandan, Shyam and Ouillon, Guy and Sornette, Didier},
  journal={Journal of Geophysical Research: Solid Earth},
  volume={124},
  number={3},
  pages={2762--2780},
  year={2019},
  publisher={Wiley Online Library}
}

@article{helmstetter2003earthquake,
  title={Is earthquake triggering driven by small earthquakes?},
  author={Helmstetter, Agnes},
  journal={Physical review letters},
  volume={91},
  number={5},
  pages={058501},
  year={2003},
  publisher={APS}
}

@article{marsan2005role,
  title={The role of small earthquakes in redistributing crustal elastic stress},
  author={Marsan, David},
  journal={Geophysical Journal International},
  volume={163},
  number={1},
  pages={141--151},
  year={2005},
  publisher={Blackwell Publishing Ltd Oxford, UK}
}

@article{mignan2014fifty,
  title={Fifty years of seismic network performance in Greece (1964--2013): spatiotemporal evolution of the completeness magnitude},
  author={Mignan, Arnaud and Chouliaras, Gerasimos},
  journal={Seismological Research Letters},
  volume={85},
  number={3},
  pages={657--667},
  year={2014},
  publisher={Seismological Society of America}
}

@article{mignan2011bayesian,
  title={Bayesian estimation of the spatially varying completeness magnitude of earthquake catalogs},
  author={Mignan, A and Werner, MJ and Wiemer, S and Chen, C-C and Wu, Y-M},
  journal={Bulletin of the Seismological Society of America},
  volume={101},
  number={3},
  pages={1371--1385},
  year={2011},
  publisher={Seismological Society of America}
}

@article{woessner2005assessing,
  title={Assessing the quality of earthquake catalogues: Estimating the magnitude of completeness and its uncertainty},
  author={Woessner, Jochen and Wiemer, Stefan},
  journal={Bulletin of the Seismological Society of America},
  volume={95},
  number={2},
  pages={684--698},
  year={2005},
  publisher={Seismological Society of America}
}

@article{mizrahi2021effect,
    author = {Mizrahi, Leila and Nandan, Shyam and Wiemer, Stefan},
    title = "{The Effect of Declustering on the Size Distribution of Mainshocks}",
    journal = {Seismological Research Letters},
    year = {2021},
    month = {02},
    issn = {0895-0695},
    doi = {10.1785/0220200231},
    url = {https://doi.org/10.1785/0220200231},
    eprint = {https://pubs.geoscienceworld.org/srl/article-pdf/doi/10.1785/0220200231/5235187/srl-2020231.1.pdf},
}

@article{nandan2016systematic,
  title={Systematic assessment of the static stress triggering hypothesis using interearthquake time statistics},
  author={Nandan, Shyam and Ouillon, Guy and Woessner, Jochen and Sornette, Didier and Wiemer, Stefan},
  journal={Journal of Geophysical Research: Solid Earth},
  volume={121},
  number={3},
  pages={1890--1909},
  year={2016},
  publisher={Wiley Online Library}
}

@article{amato2008performance,
  title={Performance of the INGV National Seismic Network from 1997 to 2007},
  author={Amato, Alessandro and Mele, F},
  journal={Annals of Geophysics},
  year={2008},
  publisher={INGV}
}

@article{hutton2010earthquake,
  title={Earthquake monitoring in southern California for seventy-seven years (1932--2008)},
  author={Hutton, Kate and Woessner, Jochen and Hauksson, Egill},
  journal={Bulletin of the Seismological Society of America},
  volume={100},
  number={2},
  pages={423--446},
  year={2010},
  publisher={Seismological Society of America}
}

@article{hainzl2008impact,
  title={Impact of earthquake rupture extensions on parameter estimations of point-process models},
  author={Hainzl, Sebastian and Christophersen, A and Enescu, B},
  journal={Bulletin of the Seismological Society of America},
  volume={98},
  number={4},
  pages={2066--2072},
  year={2008},
  publisher={Seismological Society of America}
}

@article{hainzl2013impact,
  title={Impact of aseismic transients on the estimation of aftershock productivity parameters},
  author={Hainzl, Sebastian and Zakharova, Olga and Marsan, David},
  journal={Bulletin of the Seismological Society of America},
  volume={103},
  number={3},
  pages={1723--1732},
  year={2013},
  publisher={Seismological Society of America}
}

@article{seif2017estimating,
  title={Estimating ETAS: The effects of truncation, missing data, and model assumptions},
  author={Seif, Stefanie and Mignan, Arnaud and Zechar, Jeremy Douglas and Werner, Maximilian Jonas and Wiemer, Stefan},
  journal={Journal of Geophysical Research: Solid Earth},
  volume={122},
  number={1},
  pages={449--469},
  year={2017},
  publisher={Wiley Online Library}
}

@article{schorlemmer2007relm,
  title={RELM testing center},
  author={Schorlemmer, Danijel and Gerstenberger, MC},
  journal={Seismological Research Letters},
  volume={78},
  number={1},
  pages={30--36},
  year={2007},
  publisher={Seismological Society of America}
}

@article{wang2010missing,
  title={Missing links in earthquake clustering models},
  author={Wang, Qi and Jackson, David D and Zhuang, Jiancang},
  journal={Geophysical Research Letters},
  volume={37},
  number={21},
  year={2010},
  publisher={Wiley Online Library}
}

@article{schoenberg2010relationship,
  title={On the relationship between lower magnitude thresholds and bias in epidemic-type aftershock sequence parameter estimates},
  author={Schoenberg, Frederic Paik and Chu, Annie and Veen, Alejandro},
  journal={Journal of Geophysical Research: Solid Earth},
  volume={115},
  number={B4},
  year={2010},
  publisher={Wiley Online Library}
}

@article{hainzl2016rate,
  title={Rate-dependent incompleteness of earthquake catalogs},
  author={Hainzl, Sebastian},
  journal={Seismological Research Letters},
  volume={87},
  number={2A},
  pages={337--344},
  year={2016},
  publisher={Seismological Society of America}
}

@article{schoenberg2013facilitated,
  title={Facilitated estimation of ETAS},
  author={Schoenberg, Frederic Paik},
  journal={Bulletin of the Seismological Society of America},
  volume={103},
  number={1},
  pages={601--605},
  year={2013},
  publisher={Seismological Society of America}
}

@article{tinti1987confidence,
  title={Confidence intervals of b values for grouped magnitudes},
  author={Tinti, Stefano and Mulargia, Francesco},
  journal={Bulletin of the Seismological Society of America},
  volume={77},
  number={6},
  pages={2125--2134},
  year={1987},
  publisher={The Seismological Society of America}
}

@article{stallone2020missing,
  title={Missing earthquake data reconstruction in the space-time-magnitude domain},
  author={Stallone, Angela and Falcone, Giuseppe},
  year={2020}
}

@article{zhuang2017data,
  title={Data completeness of the Kumamoto earthquake sequence in the JMA catalog and its influence on the estimation of the ETAS parameters},
  author={Zhuang, Jiancang and Ogata, Yosihiko and Wang, Ting},
  journal={Earth, Planets and Space},
  volume={69},
  number={1},
  pages={36},
  year={2017},
  publisher={Springer}
}

@article{mignan2012estimating,
  title={Estimating the magnitude of completeness for earthquake catalogs},
  author={Mignan, Arnaud and Woessner, Jochen},
  journal={Community Online Resource for Statistical Seismicity Analysis},
  pages={1--45},
  year={2012},
  publisher={Community Online Resource for Statistical Seismicity Analysis}
}

@article{schorlemmer2008probability,
  title={Probability of detecting an earthquake},
  author={Schorlemmer, Danijel and Woessner, Jochen},
  journal={Bulletin of the Seismological Society of America},
  volume={98},
  number={5},
  pages={2103--2117},
  year={2008},
  publisher={Seismological Society of America}
}

@article{sornette2005apparent,
  title={Apparent clustering and apparent background earthquakes biased by undetected seismicity},
  author={Sornette, Didier and Werner, Maximillian J},
  journal={Journal of Geophysical Research: Solid Earth},
  volume={110},
  number={B9},
  year={2005},
  publisher={Wiley Online Library}
}

@article{sornette2005constraints,
  title={Constraints on the size of the smallest triggering earthquake from the epidemic-type aftershock sequence model, B{\aa}th's law, and observed aftershock sequences},
  author={Sornette, Didier and Werner, Maximilian J},
  journal={Journal of Geophysical Research: Solid Earth},
  volume={110},
  number={B8},
  year={2005},
  publisher={Wiley Online Library}
}

@article{helmstetter2006comparison,
  title={Comparison of short-term and time-independent earthquake forecast models for southern California},
  author={Helmstetter, Agnes and Kagan, Yan Y and Jackson, David D},
  journal={Bulletin of the Seismological Society of America},
  volume={96},
  number={1},
  pages={90--106},
  year={2006},
  publisher={Seismological Society of America}
}

@article{strader2017prospective,
  title={Prospective and retrospective evaluation of five-year earthquake forecast models for California},
  author={Strader, Anne and Schneider, Max and Schorlemmer, Danijel},
  journal={Geophysical Journal International},
  volume={211},
  number={1},
  pages={239--251},
  year={2017},
  publisher={Oxford University Press}
}

@article{helmstetter2005importance,
  title={Importance of small earthquakes for stress transfers and earthquake triggering},
  author={Helmstetter, Agn{\`e}s and Kagan, Yan Y and Jackson, David D},
  journal={Journal of Geophysical Research: Solid Earth},
  volume={110},
  number={B5},
  year={2005},
  publisher={Wiley Online Library}
}

@article{wiemer2000minimum,
  title={Minimum magnitude of completeness in earthquake catalogs: Examples from Alaska, the western United States, and Japan},
  author={Wiemer, Stefan and Wyss, Max},
  journal={Bulletin of the Seismological Society of America},
  volume={90},
  number={4},
  pages={859--869},
  year={2000},
  publisher={Seismological Society of America}
}

@article{kagan2004short,
  title={Short-term properties of earthquake catalogs and models of earthquake source},
  author={Kagan, Yan Y},
  journal={Bulletin of the Seismological Society of America},
  volume={94},
  number={4},
  pages={1207--1228},
  year={2004},
  publisher={Seismological Society of America}
}

@article{felzer2007appendix,
  title={Appendix I: calculating California seismicity rates},
  author={Felzer, Karen R},
  journal={US Geol. Surv. Open-File Rept. 2007-1437I},
  year={2007}
}

@article{nandan2020earth,
  title={Is the Earth crust operating at a critical point?},
  author={Nandan, Shyam and Ram, Sumit Kumar and Ouillon, Guy and Sornette, Didier},
  journal={arXiv preprint arXiv:2012.06013},
  year={2020}
}

@article{hainzl2016apparent,
  title={Apparent triggering function of aftershocks resulting from rate-dependent incompleteness of earthquake catalogs},
  author={Hainzl, Sebastian},
  journal={Journal of Geophysical Research: Solid Earth},
  volume={121},
  number={9},
  pages={6499--6509},
  year={2016},
  publisher={Wiley Online Library}
}

@article{de2018overlap,
  title={The overlap of aftershock coda waves and short-term postseismic forecasting},
  author={De Arcangelis, L and Godano, C and Lippiello, E},
  journal={Journal of Geophysical Research: Solid Earth},
  volume={123},
  number={7},
  pages={5661--5674},
  year={2018},
  publisher={Wiley Online Library}
}

@article{lippiello2007dynamical,
  title={Dynamical scaling and generalized Omori law},
  author={Lippiello, Eugenio and Bottiglieri, M and Godano, Cataldo and de Arcangelis, Lucilla},
  journal={Geophysical Research Letters},
  volume={34},
  number={23},
  year={2007},
  publisher={Wiley Online Library}
}

@article{gerstenberger2020probabilistic,
  title={Probabilistic seismic hazard analysis at regional and national scales: State of the art and future challenges},
  author={Gerstenberger, Matthew C and Marzocchi, Warner and Allen, Trevor and Pagani, Marco and Adams, Janice and Danciu, Laurentiu and Field, Edward H and Fujiwara, H and Luco, Nicolas and Ma, K-F and others},
  journal={Reviews of Geophysics},
  volume={58},
  number={2},
  pages={e2019RG000653},
  year={2020},
  publisher={Wiley Online Library}
}

@article{narteau2009common,
  title={Common dependence on stress for the two fundamental laws of statistical seismology},
  author={Narteau, Cl{\'e}ment and Byrdina, Svetlana and Shebalin, Peter and Schorlemmer, Danijel},
  journal={Nature},
  volume={462},
  number={7273},
  pages={642--645},
  year={2009},
  publisher={Nature Publishing Group}
}

@article{lolli2006comparing,
  title={Comparing different models of aftershock rate decay: The role of catalog incompleteness in the first times after main shock},
  author={Lolli, Barbara and Gasperini, Paolo},
  journal={Tectonophysics},
  volume={423},
  number={1-4},
  pages={43--59},
  year={2006},
  publisher={Elsevier}
}

@article{shcherbakov2004generalized,
  title={A generalized Omori's law for earthquake aftershock decay},
  author={Shcherbakov, Robert and Turcotte, Donald L and Rundle, John B},
  journal={Geophysical research letters},
  volume={31},
  number={11},
  year={2004},
  publisher={Wiley Online Library}
}

@article{cao2002temporal,
  title={Temporal variation of seismic b-values beneath northeastern Japan island arc},
  author={Cao, Aimin and Gao, Stephen S},
  journal={Geophysical research letters},
  volume={29},
  number={9},
  pages={48--1},
  year={2002},
  publisher={Wiley Online Library}
}

@article{amorese2007applying,
  title={Applying a change-point detection method on frequency-magnitude distributions},
  author={Amorese, Daniel},
  journal={Bulletin of the Seismological Society of America},
  volume={97},
  number={5},
  pages={1742--1749},
  year={2007},
  publisher={Seismological Society of America}
}

@article{rydelek1989testing,
  title={Testing the completeness of earthquake catalogues and the hypothesis of self-similarity},
  author={Rydelek, Paul A and Sacks, I Selwyn},
  journal={Nature},
  volume={337},
  number={6204},
  pages={251--253},
  year={1989},
  publisher={Nature Publishing Group}
}

@article{nanjo2010analysis,
  title={Analysis of the completeness magnitude and seismic network coverage of Japan},
  author={Nanjo, KZ and Ishibe, T and Tsuruoka, H and Schorlemmer, D and Ishigaki, Y and Hirata, N},
  journal={Bulletin of the Seismological Society of America},
  volume={100},
  number={6},
  pages={3261--3268},
  year={2010},
  publisher={Seismological Society of America}
}

@article{omi2014estimating,
  title={Estimating the ETAS model from an early aftershock sequence},
  author={Omi, Takahiro and Ogata, Yosihiko and Hirata, Yoshito and Aihara, Kazuyuki},
  journal={Geophysical Research Letters},
  volume={41},
  number={3},
  pages={850--857},
  year={2014},
  publisher={Wiley Online Library}
}

@article{llenos2020regionally,
  title={Regionally Optimized Background Earthquake Rates from ETAS (ROBERE) for probabilistic seismic hazard assessment},
  author={Llenos, Andrea L and Michael, Andrew J},
  journal={Bulletin of the Seismological Society of America},
  volume={110},
  number={3},
  pages={1172--1190},
  year={2020},
  publisher={Seismological Society of America}
}

@article{nandan2021triggering,
  title={Triggering of large earthquakes is driven by their twins},
  author={Nandan, Shyam and Ouillon, Guy and Sornette, Didier},
  journal={arXiv preprint arXiv:2104.04592},
  year={2021}
}

@article{clauset2009power,
  title={Power-law distributions in empirical data},
  author={Clauset, Aaron and Shalizi, Cosma Rohilla and Newman, Mark EJ},
  journal={SIAM review},
  volume={51},
  number={4},
  pages={661--703},
  year={2009},
  publisher={SIAM}
}

@article{anss,
    title={Advanced National Seismic System (ANSS) Comprehensive Catalog of Earthquake Events and Products: Various},
    year={2017},
    author={{U.S. Geological Survey, Earthquake Hazards Program}},
    doi = {10.5066/F7MS3QZH},
    url = {https://doi.org/10.5066/F7MS3QZH}
}

@article{helmstetter2014adaptive,
  title={Adaptive smoothing of seismicity in time, space, and magnitude for time-dependent earthquake forecasts for California},
  author={Helmstetter, Agn{\`e}s and Werner, Maximilian J},
  journal={Bulletin of the Seismological Society of America},
  volume={104},
  number={2},
  pages={809--822},
  year={2014},
  publisher={Seismological Society of America}
}

@article{ogata2006immediate,
  title={Immediate and updated forecasting of aftershock hazard},
  author={Ogata, Yosihiko and Katsura, Koichi},
  journal={Geophysical research letters},
  volume={33},
  number={10},
  year={2006},
  publisher={Wiley Online Library}
}

@article{guo1997statistical,
  title={Statistical relations between the parameters of aftershocks in time, space, and magnitude},
  author={Guo, Zhenqi and Ogata, Yosihiko},
  journal={Journal of Geophysical Research: Solid Earth},
  volume={102},
  number={B2},
  pages={2857--2873},
  year={1997},
  publisher={Wiley Online Library}
}

@article{taroni2018prospective,
  title={Prospective CSEP evaluation of 1-day, 3-month, and 5-yr earthquake forecasts for Italy},
  author={Taroni, Matteo and Marzocchi, Warner and Schorlemmer, Danijel and Werner, Maximilian Jonas and Wiemer, Stefan and Zechar, Jeremy Douglas and Heiniger, Lukas and Euchner, Fabian},
  journal={Seismological Research Letters},
  volume={89},
  number={4},
  pages={1251--1261},
  year={2018},
  publisher={Seismological Society of America}
}

@article{woessner2011retrospective,
  title={A retrospective comparative forecast test on the 1992 Landers sequence},
  author={Woessner, J and Hainzl, Sebastian and Marzocchi, W and Werner, MJ and Lombardi, AM and Catalli, F and Enescu, B and Cocco, M and Gerstenberger, MC and Wiemer, S},
  journal={Journal of Geophysical Research: Solid Earth},
  volume={116},
  number={B5},
  year={2011},
  publisher={Wiley Online Library}
}

@article{cattania2018forecasting,
  title={The forecasting skill of physics-based seismicity models during the 2010--2012 Canterbury, New Zealand, earthquake sequence},
  author={Cattania, Camilla and Werner, Maximilian J and Marzocchi, Warner and Hainzl, Sebastian and Rhoades, David and Gerstenberger, Matthew and Liukis, Maria and Savran, William and Christophersen, Annemarie and Helmstetter, Agn{\`e}s and others},
  journal={Seismological Research Letters},
  volume={89},
  number={4},
  pages={1238--1250},
  year={2018},
  publisher={Seismological Society of America}
}

@article{mancini2019improving,
  title={Improving physics-based aftershock forecasts during the 2016--2017 Central Italy Earthquake Cascade},
  author={Mancini, S and Segou, M and Werner, MJ and Cattania, C},
  journal={Journal of Geophysical Research: Solid Earth},
  volume={124},
  number={8},
  pages={8626--8643},
  year={2019},
  publisher={Wiley Online Library}
}

@article{mancini2020predictive,
  title={The predictive skills of elastic Coulomb rate-and-state aftershock forecasts during the 2019 Ridgecrest, California, earthquake sequence},
  author={Mancini, Simone and Segou, Margarita and Werner, Maximilian Jonas and Parsons, Tom},
  journal={Bulletin of the Seismological Society of America},
  volume={110},
  number={4},
  pages={1736--1751},
  year={2020},
  publisher={Seismological Society of America}
}

@article{field2017synoptic,
  title={A synoptic view of the third Uniform California Earthquake Rupture Forecast (UCERF3)},
  author={Field, Edward H and Jordan, Thomas H and Page, Morgan T and Milner, Kevin R and Shaw, Bruce E and Dawson, Timothy E and Biasi, Glenn P and Parsons, Tom and Hardebeck, Jeanne L and Michael, Andrew J and others},
  journal={Seismological Research Letters},
  volume={88},
  number={5},
  pages={1259--1267},
  year={2017},
  publisher={Seismological Society of America}
}

@article{marzocchi2014establishment,
  title={The establishment of an operational earthquake forecasting system in Italy},
  author={Marzocchi, Warner and Lombardi, Anna Maria and Casarotti, Emanuele},
  journal={Seismological Research Letters},
  volume={85},
  number={5},
  pages={961--969},
  year={2014},
  publisher={Seismological Society of America}
}

@article{rhoades2016retrospective,
  title={Retrospective tests of hybrid operational earthquake forecasting models for Canterbury},
  author={Rhoades, DA and Liukis, M and Christophersen, A and Gerstenberger, MC},
  journal={Geophysical Journal International},
  volume={204},
  number={1},
  pages={440--456},
  year={2016},
  publisher={Oxford University Press}
}

@article{bray2014voronoi,
  title={Voronoi residual analysis of spatial point process models with applications to California earthquake forecasts},
  author={Bray, Andrew and Wong, Ka and Barr, Christopher D and Schoenberg, Frederic Paik},
  journal={The Annals of Applied Statistics},
  volume={8},
  number={4},
  pages={2247--2267},
  year={2014},
  publisher={Institute of Mathematical Statistics}
}

@article{gordon2021nonparametric,
  title={A nonparametric Hawkes model for forecasting California seismicity},
  author={Gordon, Joshua Seth and Fox, Eric Warren and Schoenberg, Frederic Paik},
  journal={Bulletin of the Seismological Society of America},
  volume={111},
  number={4},
  pages={2216--2234},
  year={2021},
  publisher={Seismological Society of America}
}

@article{clements2011residual,
  title={Residual analysis methods for space-time point processes with applications to earthquake forecast models in California},
  author={Clements, Robert Alan and Schoenberg, Frederic Paik and Schorlemmer, Danijel},
  journal={The Annals of applied statistics},
  pages={2549--2571},
  year={2011},
  publisher={JSTOR}
}

@article{harte2005entropy,
  title={The entropy score and its uses in earthquake forecasting},
  author={Harte, David and Vere-Jones, David},
  journal={Pure and Applied Geophysics},
  volume={162},
  number={6},
  pages={1229--1253},
  year={2005},
  publisher={Springer}
}

@article{zechar2013regional,
  title={Regional earthquake likelihood models I: First-order results},
  author={Zechar, J Douglas and Schorlemmer, Danijel and Werner, Maximilian J and Gerstenberger, Matthew C and Rhoades, David A and Jordan, Thomas H},
  journal={Bulletin of the Seismological Society of America},
  volume={103},
  number={2A},
  pages={787--798},
  year={2013},
  publisher={Seismological Society of America}
}

@article{rhoades2011efficient,
  title={Efficient testing of earthquake forecasting models},
  author={Rhoades, David A and Schorlemmer, Danijel and Gerstenberger, Matthew C and Christophersen, Annemarie and Zechar, J Douglas and Imoto, Masajiro},
  journal={Acta Geophysica},
  volume={59},
  number={4},
  pages={728--747},
  year={2011},
  publisher={Springer}
}

@article{gordon2015voronoi,
  title={Voronoi residuals and other residual analyses applied to CSEP earthquake forecasts},
  author={Gordon, Joshua Seth and Clements, Robert Alan and Schoenberg, Frederic Paik and Schorlemmer, Danijel},
  journal={Spatial Statistics},
  volume={14},
  pages={133--150},
  year={2015},
  publisher={Elsevier}
}

@article{nichols2014assessing,
  title={Assessing the dependency between the magnitudes of earthquakes and the magnitudes of their aftershocks},
  author={Nichols, Kevin and Schoenberg, Frederic Paik},
  journal={Environmetrics},
  volume={25},
  number={3},
  pages={143--151},
  year={2014},
  publisher={Wiley Online Library}
}

@article{spassiani2016exploring,
  title={Exploring the relationship between the magnitudes of seismic events},
  author={Spassiani, Ilaria and Sebastiani, Giovanni},
  journal={Journal of Geophysical Research: Solid Earth},
  volume={121},
  number={2},
  pages={903--916},
  year={2016},
  publisher={Wiley Online Library}
}

@article{nandan2021global,
  title={Global models for short-term earthquake forecasting and predictive skill assessment},
  author={Nandan, Shyam and Kamer, Yavor and Ouillon, Guy and Hiemer, Stefan and Sornette, Didier},
  journal={The European Physical Journal Special Topics},
  volume={230},
  number={1},
  pages={425--449},
  year={2021},
  publisher={Springer}
}

@article{ouillon2005magnitude,
  title={Magnitude-dependent Omori law: Theory and empirical study},
  author={Ouillon, Guy and Sornette, Didier},
  journal={Journal of Geophysical Research: Solid Earth},
  volume={110},
  number={B4},
  year={2005},
  publisher={Wiley Online Library}
}

@article{sornette2005multifractal,
  title={Multifractal scaling of thermally activated rupture processes},
  author={Sornette, D and Ouillon, G},
  journal={Physical review letters},
  volume={94},
  number={3},
  pages={038501},
  year={2005},
  publisher={APS}
}

@article{guo2015improved,
  title={An improved space-time ETAS model for inverting the rupture geometry from seismicity triggering},
  author={Guo, Yicun and Zhuang, Jiancang and Zhou, Shiyong},
  journal={Journal of Geophysical Research: Solid Earth},
  volume={120},
  number={5},
  pages={3309--3323},
  year={2015},
  publisher={Wiley Online Library}
}

@article{wells1994new,
  title={New empirical relationships among magnitude, rupture length, rupture width, rupture area, and surface displacement},
  author={Wells, Donald L and Coppersmith, Kevin J},
  journal={Bulletin of the seismological Society of America},
  volume={84},
  number={4},
  pages={974--1002},
  year={1994},
  publisher={The Seismological Society of America}
}

@article{meier2014search,
  title={A search for evidence of secondary static stress triggering during the 1992 Mw7. 3 Landers, California, earthquake sequence},
  author={Meier, M-A and Werner, MJ and Woessner, J and Wiemer, S},
  journal={Journal of Geophysical Research: Solid Earth},
  volume={119},
  number={4},
  pages={3354--3370},
  year={2014},
  publisher={Wiley Online Library}
}

@article{segou2013comparative,
  title={Comparative evaluation of physics-based and statistical forecasts in northern California},
  author={Segou, M and Parsons, T and Ellsworth, W},
  journal={Journal of Geophysical Research: Solid Earth},
  volume={118},
  number={12},
  pages={6219--6240},
  year={2013},
  publisher={Wiley Online Library}
}

@article{savran2020pseudoprospective,
  title={Pseudoprospective Evaluation of UCERF3-ETAS Forecasts during the 2019 Ridgecrest Sequence},
  author={Savran, William H and Werner, Maximilian J and Marzocchi, Warner and Rhoades, David A and Jackson, David D and Milner, Kevin and Field, Edward and Michael, Andrew},
  journal={Bulletin of the Seismological Society of America},
  volume={110},
  number={4},
  pages={1799--1817},
  year={2020},
  publisher={Seismological Society of America}
}

@article{baath1965lateral,
  title={Lateral inhomogeneities of the upper mantle},
  author={B{\aa}th, Markus},
  journal={Tectonophysics},
  volume={2},
  number={6},
  pages={483--514},
  year={1965},
  publisher={Elsevier}
}

@article{utsu1961statistical,
  title={A statistical study on the occurrence of aftershocks},
  author={Utsu, Tokuji},
  journal={Geophys. Mag.},
  volume={30},
  pages={521--605},
  year={1961}
}

\clearpage
\section{Tables}
        
    \begin{table}[ht]
        \centering
        \caption{ETAS and PETAI parameters inferred for California. First column shows parameters when constant $m_c$ of 3.1 is assumed. Second and third column show parameters when time-varying $m_c$ is accounted for, and fourth and fifth column show parameters when PETAI inversion is applied. Note that the originally derived parameters are given in Columns 1, 2, and 4. Columns 3 and 5 show the parameters of Columns 2 and 4, transformed (as described in Equations \ref{eq:transform_d} - \ref{eq:transform_mu}) to a reference magnitude of 3.1 to allow comparison with Column 1. Productivity exponent $\alpha = a - \rho \gamma$ and branching ratio $\eta$ are not directly inverted but inferred from the inverted parameters.}
        \begin{tabular}{l|r|rr|rr}
        \label{tab:params_cali}
        parameter         & \multicolumn{1}{c|}{$m_c\equiv$ const.} & \multicolumn{2}{c|}{$m_c(t)$} & \multicolumn{2}{c}{$f(m, t)$} \\ \hline
        $m_{ref}$         & 3.1                                     & 2.4           & 3.1           & 2.5            & 3.1           \\
        $\log_{10}(\mu)$  & -6.86                                   & -5.97         & -6.68         & -6.35          & -6.97         \\
        $\log_{10}(k_0)$  & -2.53                                   & -2.63         & -2.36         & -2.70          & -2.49         \\
        $a$               & 1.74                                    & 1.86          & 1.86          & 1.92           & 1.92          \\
        $\log_{10}(c)$    & -2.97                                   & -2.52         & -2.52         & -2.85          & -2.85         \\
        $\omega$          & -0.05                                   & -0.02         & -0.02         & -0.06          & -0.06         \\
        $\log_{10}(\tau)$ & 4.03                                    & 3.57          & 3.57          & 3.92           & 3.92          \\
        $\log_{10}(d)$    & -0.51                                   & -0.86         & -0.45         & -0.76          & -0.45         \\
        $\gamma$          & 1.19                                    & 1.35          & 1.35          & 1.22           & 1.22          \\
        $\rho$            & 0.60                                    & 0.67          & 0.67          & 0.67           & 0.67          \\
        $\log_{10}(t_R)$  & n/a                                     & n/a           & n/a           & -2.86          & -2.86         \\
        $\beta$           & 2.33                                    & 2.32          & 2.32          & 2.37           & 2.37      \\ \hline
        $a - \rho \gamma$ & 1.03                                    & 0.95          & 0.95          & 1.09           & 1.09          \\
        $\eta$            & 0.94                                    & 0.95          & 0.95          & 0.93           & 0.93   
        \end{tabular}
    \end{table}
    
\clearpage
\section{Figures}
    
    \begin{figure}

        \centering

\tikzstyle{startstop} = [rectangle, rounded corners, minimum width=3cm, minimum height=1cm,text centered, text width=3cm, draw=black, fill=red!30]
\tikzstyle{io} = [trapezium, trapezium left angle=70, trapezium right angle=110, minimum width=3cm, minimum height=1cm, text centered, text width=3cm, draw=black, fill=blue!30]
\tikzstyle{process} = [rectangle, minimum width=3cm, minimum height=1cm, text centered, text width=3cm, draw=black, fill=orange!30]
\tikzstyle{decision} = [diamond, minimum width=3cm, minimum height=1cm, text centered, text width=2.2cm, draw=black, fill=green!30]
\tikzstyle{text} = [rectangle, minimum width=1cm, minimum height=1cm]
\tikzstyle{arrow} = [thick,->,>=stealth]

\resizebox{\textwidth}{!}{%
    \begin{tikzpicture}[node distance=2cm]

        \node (start) [startstop] {start};
        \node (mcconst) [io, below of=start] {define\\$m_c \equiv$ const.};
        \node (pro1) [process, below of=mcconst] {
            estimate initial ETAS parameters ($\mathcal{E}$)
            ($\rightarrow$ Section \ref{sec:zeroone_etas})
        };
        \node (etasconst) [io, below of=pro1] {
            initial\\ETAS parameters\\$(\mathcal{E})$
        };
        \node (pro3) [process, below of =etasconst, , yshift=-3cm] {
            estimate ETAS parameters ($\mathcal{E}$)
            with incomplete detection\\
            ($\rightarrow$ Section \ref{sec:pdet_etas})
        };
        \coordinate (left pro3) at ($(pro3)-(3.0,-2.5cm)$);
        \node (etasnew) [io, below of=pro3, yshift=-0.5cm] {
            newly estimated\\ETAS parameters\\$(\mathcal{E})$
        };
        \node (dec1) [decision, below of=etasnew, yshift=-1.0cm] {
            $\mathcal{E}$\\converged*?
        };
        \node (allnew) [io, below of=dec1, yshift=-1.3cm] {
            newly estimated\\ETAS parameters\\$(\mathcal{E})$\\
            and incompleteness\\$(\mathcal{I} = \lambda_i, t_R, \beta)$
        };
        \node (stop) [startstop, below of=allnew, yshift=-0.3cm] {end};
        \draw [arrow] (start) -- (mcconst);
        \draw [arrow] (mcconst) -- (pro1);
        \draw [arrow] (pro1) -- (etasconst);
        \draw [arrow] (pro3) -- (etasnew);
        \draw [arrow] (etasnew) -- (dec1);
        \draw [arrow] (dec1) -- node[anchor=east] {yes} (allnew);
        \draw [arrow] (allnew) -- (stop);

        \node (start2) [startstop, right of=start, xshift=5cm] {start: estimate incompleteness\\($\mathcal{I} = \lambda_i, t_R, \beta$)};
        \coordinate (left start2) at ($(start2)-(4.0,0.0cm)$);
        \coordinate (left start2b) at ($(start2)-(4.5,0.0cm)$);
        \node (lambda0) [io, below of=start2] {define\\$\lambda_i = 0 \quad \forall i$};
        \node (trb) [process, below of=lambda0] {
            estimate \\initial ($t_R, \beta$)\\
            ($\rightarrow$ Section \ref{sec:tr_est})
        };
        \coordinate (left trb) at ($(trb)-(3.0,0.0cm)$);
        \node (trbnew) [io, below of=trb] {initial\\ ($t_R, \beta$)};
        
        \node (trb2) [process, below of=trbnew, yshift=-3cm] {
            estimate ($t_R, \beta$)
            ($\rightarrow$ Section \ref{sec:tr_est})
        };
        \coordinate (left trb2) at ($(trb2)-(3.0,-2.5cm)$);
        \node (trbnew2) [io, right of=etasnew, xshift=5cm] {newly estimated\\ ($t_R, \beta$)};
        \node (dec2) [decision, right of=dec1, xshift=5cm] {
            ($t_R, \beta$)\\converged**? 
        };
        \node (newtrlam) [io, right of=allnew, xshift=5cm] {
            newly estimated incompleteness\\$(\mathcal{I} = \lambda_i, t_R, \beta)$
        };
        \node (stop2) [startstop, right of=stop, xshift=5cm] {end incompleteness estimation};
        \coordinate (left stop2) at ($(stop2)-(4.0,0.0cm)$);
        
        \draw [arrow] (etasconst) -| (left start2b) -- (start2);
        \draw [arrow] (stop2) -- (left stop2) |- (pro3);
        \draw [arrow] (dec1) -| node[anchor=north west] {no} (left pro3) -| (left start2) --  (start2);
        
        \draw [arrow] (start2) -- (lambda0);
        \draw [arrow] (lambda0) -- (trb);
        \draw [arrow] (trb) -- (trbnew);
        \draw [arrow] (trb2) -- (trbnew2);
        \draw [arrow] (trbnew2) -- (dec2);
        \draw [arrow] (dec2) -- node[anchor=east] {yes} (newtrlam);
        \draw [arrow] (newtrlam) -- (stop2);

        \node (start3) [startstop, right of=start2, xshift=5cm] {start:\\estimate rates $\lambda_i$};
        \coordinate (left start3) at ($(start3)-(4.0,0.0cm)$);
        \coordinate (left start3b) at ($(start3)-(4.5,0.0cm)$);
        \node (xi0) [process, below of=start3] {
            calculate $\xi(t_i)$
            ($\rightarrow$ Equation \ref{eq:xi})
        };
        \coordinate (left xi0) at ($(xi0)-(3.0,0.0cm)$);
        \node (xinew) [io, below of=xi0] {newly calculated\\ $\xi(t_i)$};
        \node (calclam) [process, below of=xinew] {
            calculate $\lambda_i$\\
            ($\rightarrow$ Equation \ref{eq:lambda_new})
        };
        \node (newlam2) [io, below of=calclam] {newly calculated\\ $\lambda_i$};
        \node (dec3) [decision, below of=newlam2, yshift=-0.7cm] {
            $\lambda_i$\\converged***?
        };
        \node (newestlam) [io, below of=dec3, yshift=-0.7cm] {
            newly estimated\\$\lambda_i$
        };
        \node (stop3) [startstop, below of=newestlam] {end rates\\estimation};
        \coordinate (left stop3) at ($(stop3)-(4.0,0.0cm)$);

        \draw [arrow] (start3) -- (xi0);
        \draw [arrow] (xi0) -- (xinew);
        \draw [arrow] (xinew) -- (calclam);
        \draw [arrow] (calclam) -- (newlam2);
        \draw [arrow] (newlam2) -- (dec3);
        \draw [arrow] (dec3) -- node[anchor=east] {yes} (newestlam);
        \draw [arrow] (dec3) -| node[anchor=north west] {no} (left xi0) -- (xi0);
        \draw [arrow] (newestlam) -- (stop3);
        
        \draw [arrow] (trbnew) -| (left start3b) -- (start3);
        \draw [arrow] (stop3) -- (left stop3) |- (trb2);
        \draw [arrow] (dec2) -| node[anchor=north west] {no} (left trb2) -| (left start3) -- (start3);

    \end{tikzpicture}
}    
        \caption{Flow diagram of PETAI inversion. 
        Caption on next page.}
        \label{fig:flow}
    \end{figure}

    \addtocounter{figure}{-1}
    \begin{figure}[t!]
        \caption{(Previous page.) Flow diagram of PETAI inversion. 
        Main algorithm starts at top left and ends at bottom left.
        The middle column describes the estimation of incompleteness ($\mathcal{I} = \lambda_i, t_R, \beta$) when ETAS parameters ($\mathcal{E}$) are given.
        Note that the estimation of $(\lambda_i)_{i=1,\dots,n}$ when ETAS parameters and $(t_R, \beta)$ are fixed requires yet another loop to obtain self-consistency, as updating $\lambda_i$ (step $\Lambda$) leads to changes in the inflation factor $1 + \xi(t_i)$, which forces one to update $(\lambda_i)_{i=1,\dots,n}$.
        This sub-sub-algorithm is visualized in the right column of the flow diagram. 
        Process boxes are linked to corresponding methods and equations described in this article.\\
        *, **, ***: Convergence is reached when the estimated values of the $k^{th}$ iteration, $\hat a_k$, lie very close to the estimated values of the previous iteration, that is, if $\sum_{a\in A} |\hat a_k - \hat a_{k-1}| \leq \theta$. Here, $A$ is the set of values that are tested for convergence, *$A=\mathcal{E}$, **$A=\{t_R, \beta\}$, ***$A=\{\lambda_i, i=1, \dots, n\}$. For convergence threshold $\theta$ we use *$\theta = 10^{-3}$, **$\theta = 10^{-12}$, ***$\theta = 1$.}
    \end{figure}

    \begin{figure}[ht]
        \centering
        \includegraphics[width=1.0\textwidth]{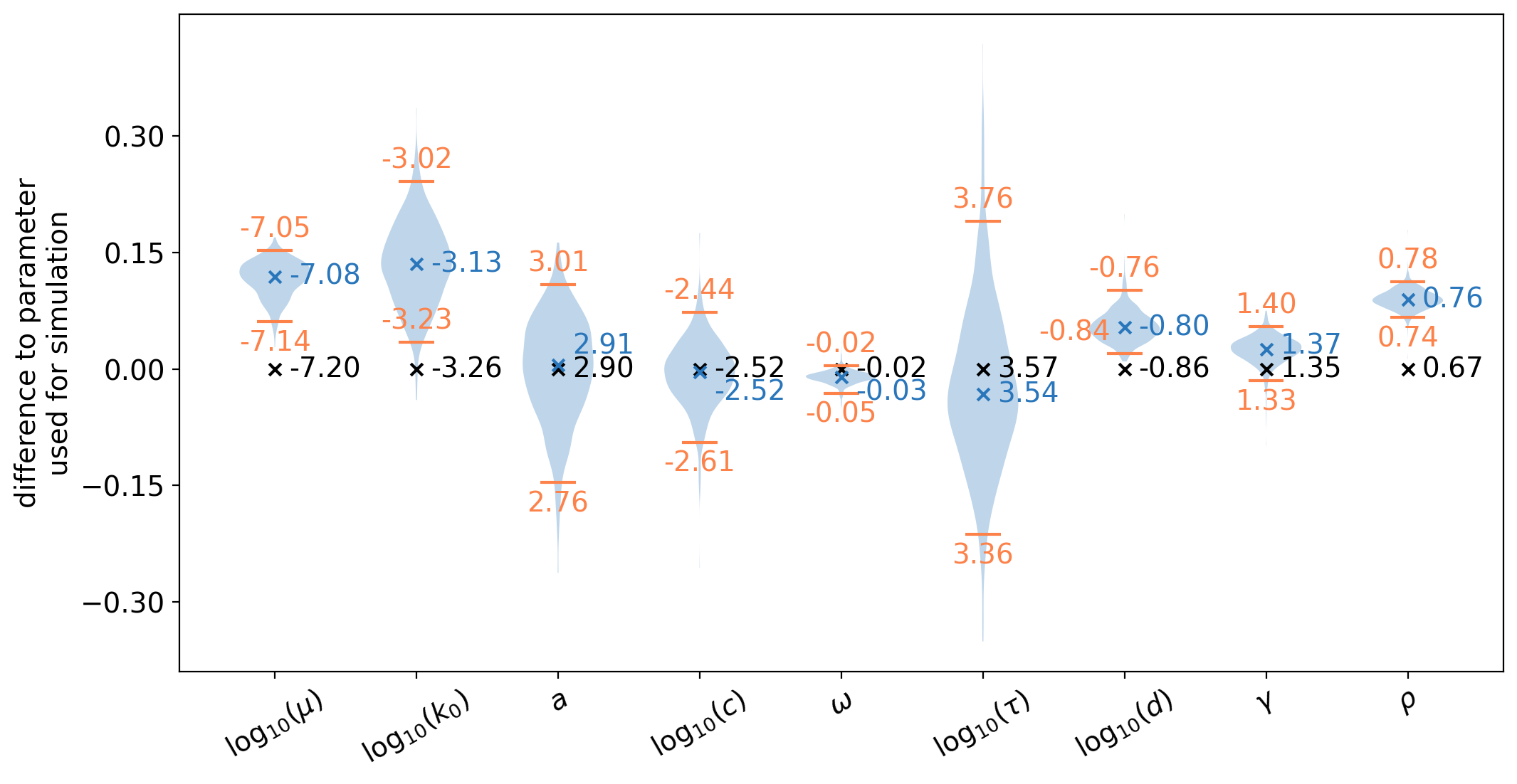}
        \caption{ETAS parameters used and inferred in synthetic test 1 (ST1). Black crosses indicate parameters used in simulation of 400 catalogs, blue crosses indicate median inverted parameters. Violins show the distribution of obtained parameters for 400 catalogs, with orange lines marking the 2.5\% and 97.5\% percentiles. Note that the y-axis gives the difference to parameters used for simulation, the actual values are written next to their marks.}
        \label{fig:hist_uncert}
    \end{figure}
    
    \begin{figure}[ht]
        \centering
        \includegraphics[width=\textwidth]{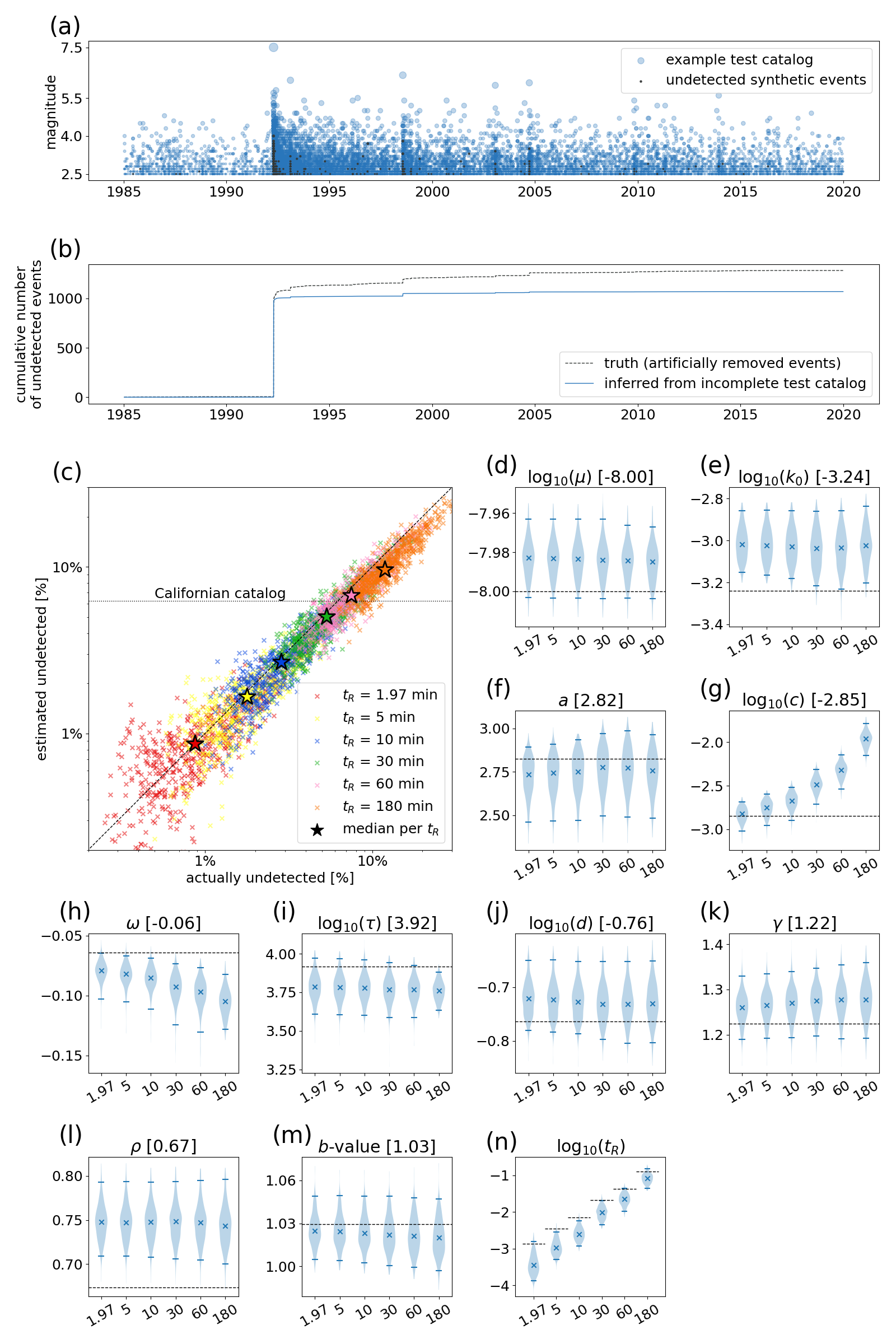}
        \caption{Results of synthetic test 2 (ST2). Caption on next page.}
        \label{fig:synth_results}
    \end{figure}
    
    \addtocounter{figure}{-1}
    \begin{figure}[t!]
        \caption{(Previous page.) Results of synthetic test 2 (ST2). (a) Earthquake magnitudes over time for one example test catalog (blue). Events marked in black were simulated, but declared as undetected. (b) Cumulative number of unobserved events over time for the catalog shown in (a). Black line marks the truth, blue line is inferred from the test catalog using PETAI. (c) Estimated fraction of undetected events versus actually removed fraction of events, for 3000 test catalogs. Different colors indicate different assumed detection efficiencies. Stars mark the median actual and estimated fraction of undetected events per $t_R$. Dashed line indicates where actual and estimated fraction coincide, dotted horizontal line indicates the estimated fraction for California. (d) - (n) ETAS and PETAI parameters inferred in synthetic test. Panel title indicates the parameter name and in square brackets the value used for simulation. Violins show the distribution of the parameter inferred from test catalogs per value of $t_R$ used, with the value of $t_R$ given on the x-axis in minutes. Crosses indicate median obtained value, blue lines indicate 95\% confidence interval, dashed line indicates the value used for simulation.}
    \end{figure}
    
    \begin{figure}[ht]
        \centering
        \includegraphics[width=\textwidth]{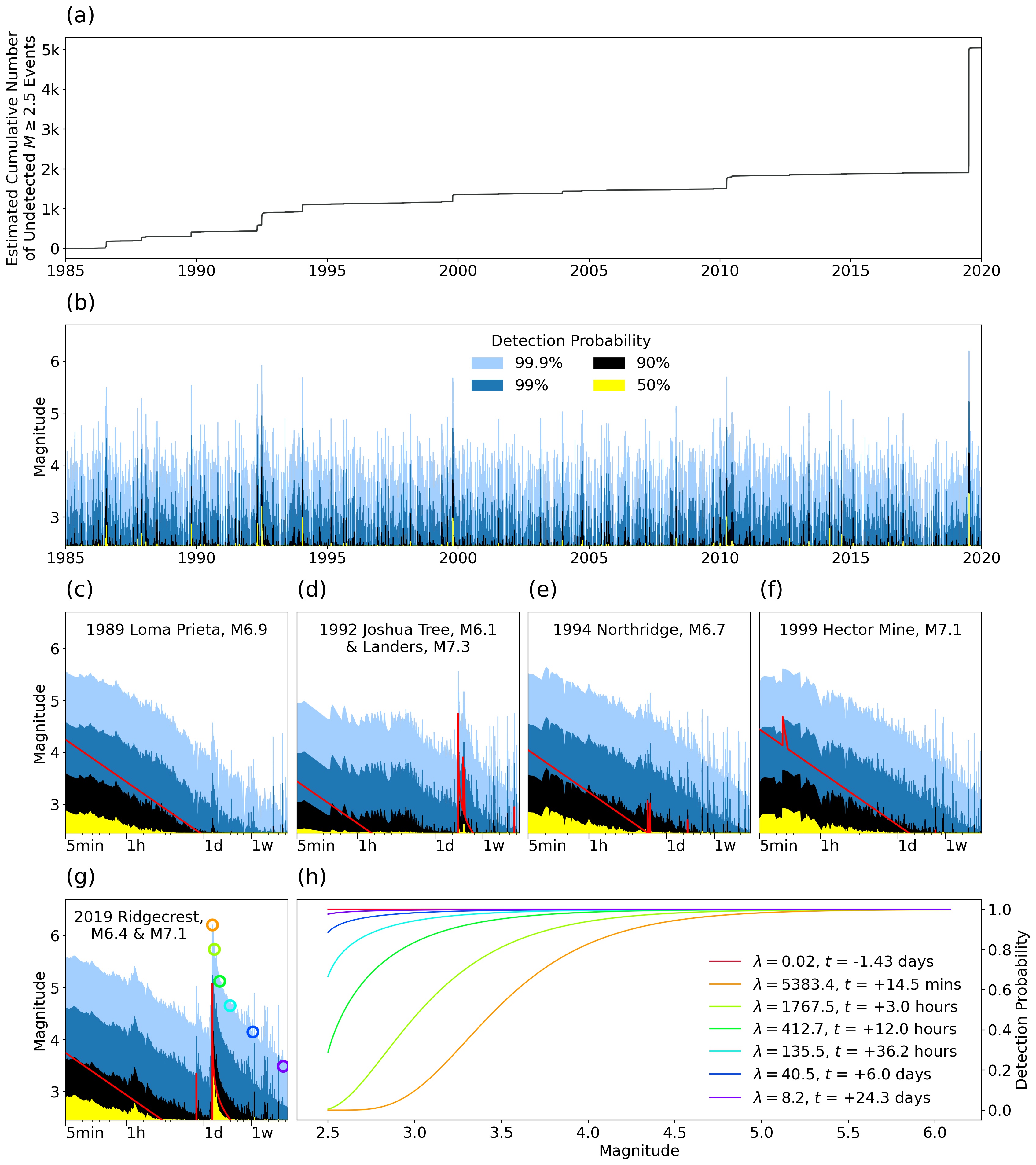}
        \caption{Aftershock Incompleteness in California. (a) Estimated cumulative number of undetected events over time. (b) Evolution of magnitude-dependent detection probability. 
        Yellow indicates a detection probability of 50\% or less.
        Black, dark blue, and light blue indicate detection probabilities of up to 90\%, 99\%, and 99.9\%, respectively.
        White area represents detection probabilities higher than 99.9\%.
        (c)-(g) Excerpts of (b) for selected large events. $x$-axes are logarithmic and show time since (first) mainshock, and range from 5 minutes to 30 days after that mainshock.
        Red lines indicate $m_c(t)$ as described by \textcite{helmstetter2006comparison}, including the effect of all $M\ge5$ events.
        Colored circles in (g) represent selected times $t_i$ and corresponding magnitude of 99.9\% detection.
        (h) Detection probability function $f(m, \lambda=\lambda(t_i))$ snapshots for the times that are highlighted in (g), plus a time prior to both mainshocks (in red).
        Time deltas are given with respect to the M7.1 mainshock. $\lambda(t_i)$ are as estimated during PETAI inversion.
        }
        \label{fig:cali_petai}
    \end{figure}
    
    \begin{figure}
        \centering
        \includegraphics[width = 1.0\textwidth]{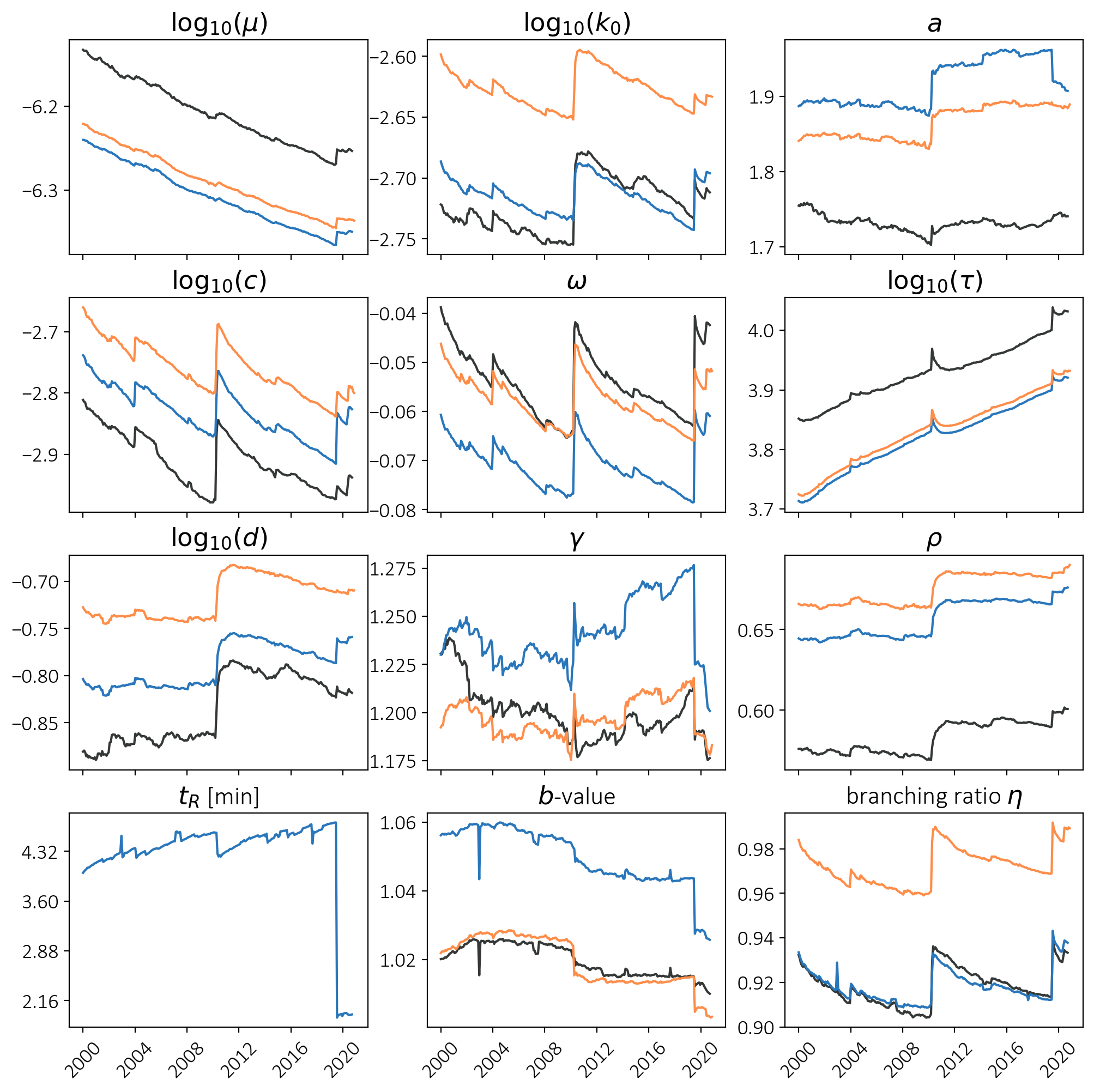}
        \caption{Evolution of ETAS and PETAI parameter estimates with increasing training catalog, when using standard inversion with $m_c = 3.1$ (black lines) or $m_c = 2.5$ (orange lines) and when using PETAI inversion (blue lines). The evolution for $t_R$ is only given for PETAI inversion because it does not exist in standard ETAS. Parameters are with respect to $m_{ref} = 2.5$, transformed using Equations \ref{eq:transform_d} - \ref{eq:transform_mu} if necessary.}
        \label{fig:par_evol_fc}
    \end{figure}

    \begin{figure}[ht]
        \centering
        \includegraphics[width=0.8\textwidth]{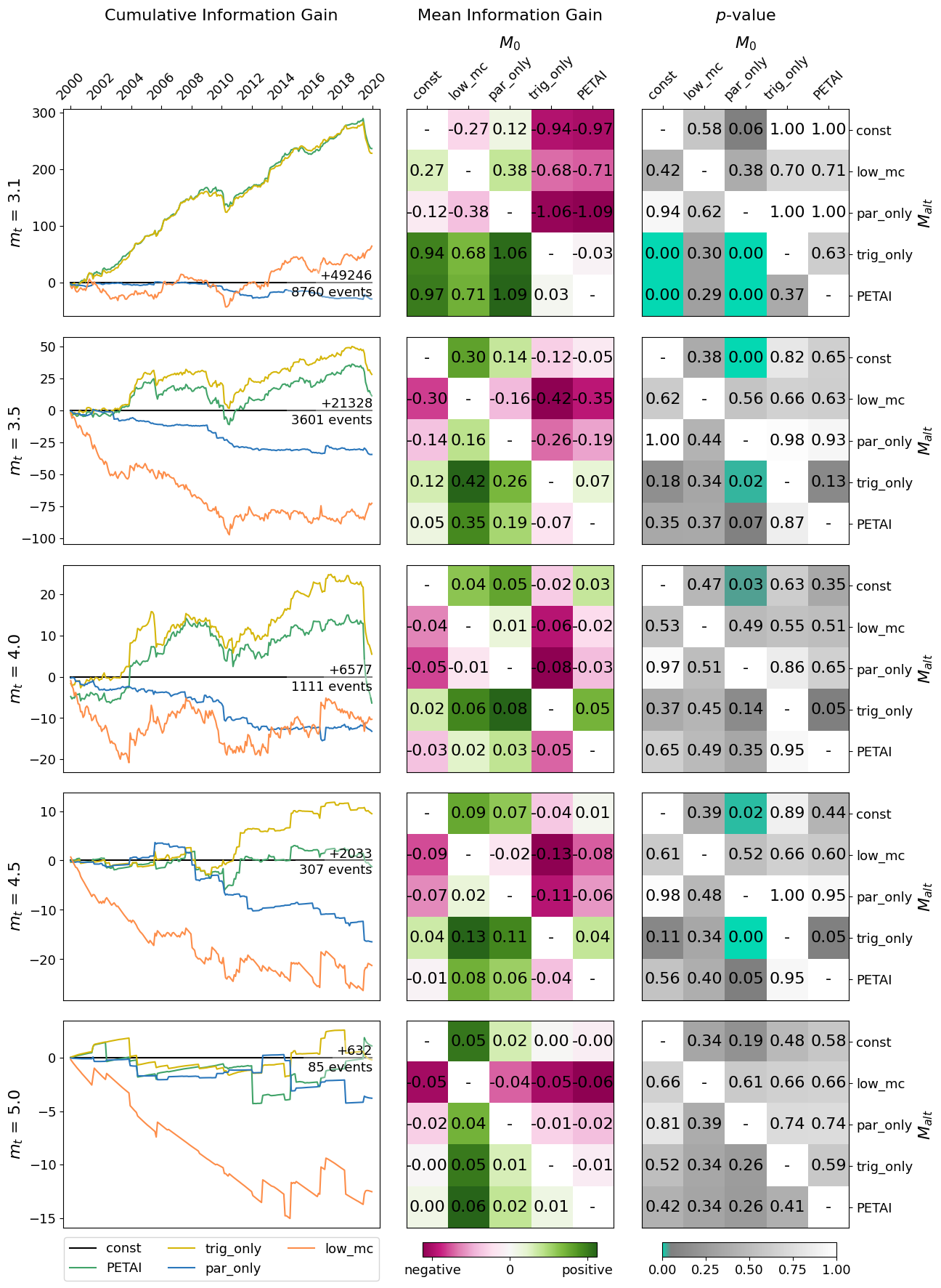}
        \caption{From left to right: Cumulative information gain for the alternative models versus the ETAS null model, mean information gain matrix, and corresponding $p$-value matrix comparing all competing models. Matrix entries represent the test of superiority of $M_{alt}$ ($y$-axis) versus $M_{0}$ ($x$-axis). From top to bottom: target magnitude thresholds $m_t$ of 3.1, 3.5, 4.0, 4.5, and 5.0. Indicated as text in the left panels is the cumulative information gain of the ETAS null model versus the STHPP model, and the number of events in all testing periods combined. Note the different $y$-axes for the left panels. Also note that the color scheme for the middle panels is different between threshold magnitudes $m_t$ and normalized with respect to the maximum absolute mean information gain for that $m_t$. Color coding for the panels on the right is such that $p$-values of 0.05 and below are green, and transition from grey to white between  0.05 and 1.}
        \label{fig:fc_results}
    \end{figure}

\end{document}


%
%


\title{Supporting Information for ``Embracing Data Incompleteness for Better Earthquake Forecasting''}
\author[1]{L. Mizrahi}
\author[1]{S. Nandan}
\author[1]{S. Wiemer}

\affil[1]{Swiss Seismological Service, ETH Zurich}
\date{}
\maketitle

%
%

%

%
%

\noindent\textbf{Contents of this file}
\begin{itemize}

\item \textbf{Text S1} Catalog simulation
\item \textbf{Text S2} Forecast evaluation
\\
\item \textbf{Figure S1} Log likelihood of observing the test data for different values of $t_R$ and $b$-value, when current rate is known

\end{itemize}


\clearpage

\renewcommand{\thesubsection}{Text S\arabic{subsection}}

\subsection{Catalog simulation}
    \label{sec:simulation}
    The following algorithm is used to simulate the continuation of a training catalog.\\
    
    Note that the synthetic catalogs referred to in Sections 4.1 and 4.2 in the article are not continuations of a training catalog, hence generation 0 (defined in the following) consists only of background events. The locations of these background events are uniformly distributed in the study region.
    Also in the case of synthetic catalog simulation, the \say{testing period}, which is referred to below, is the period for which one wishes to simulate a catalog.
    Where different models are mentioned, the base model is used for synthetic catalog simulation.

    \begin{enumerate}
        \item Background events are simulated for the testing period.
            \begin{itemize}
                \item Number of background events is drawn from a Poisson distribution with mean as given by the ETAS background rate.
                \item Occurrence times are drawn from a uniform distribution within the testing period.
                \item Locations are drawn from the locations of events in the training catalog, weighted by their probability of being background events. The locations are then randomly displaced by a distance drawn from a normal distribution with mean 0 and standard deviation of 0.1$\degree$.
                \item Magnitudes are drawn from a GR law with exponent $\beta$ as estimated in the PETAI inversion (for PETAI and trig\_only).
                For the base model and par\_only, we use the $\beta$ estimate obtained when using the formula proposed by \textcite{tinti1987confidence} for binned magnitude values, using magnitudes $M \ge 3.1$ in the training catalog.
            \end{itemize}
        \item The training catalog together with the simulated background events make up generation 0. $i_{gen} \defeq 0$.
        \item Expected number of aftershocks is calculated for all events of generation 0. In the case of the PETAI and the trig\_only model, the average number of aftershocks triggered by any event $e_i$ in the training catalog is inflated by $1+\xi(t_i)$.
        \item \label{it:simloop} Actual number of aftershocks of each event is randomly drawn from a Poisson distribution with mean as calculated in the previous step.
        \item Aftershocks of the current generation $i_{gen}$ are simulated.
            \begin{itemize}
                \item Aftershock time distance to its parent event is randomly generated according to the estimated ETAS time kernel. If aftershock time falls out of the testing period, this aftershock is discarded.
                \item Aftershock spatial distance to its parent event is randomly generated according to the estimated isotropic ETAS spatial kernel. If aftershock location falls out of the considered polygon, this aftershock is discarded.
                \item Aftershock magnitude is generated according to the GR law with exponent $\beta$ (same as for the background events).
            \end{itemize}
        \item The newly generated aftershocks now make up the next generation $i_{gen}+1$. 
        \item We move on to the next generation. $i_{gen} \defeq i_{gen} + 1$
        \item Expected number of aftershocks is calculated for all events of generation $i_{gen}$. Continue with step \ref{it:simloop}.
    \end{enumerate}
    The algorithm terminates when no aftershocks fall into the testing period anymore, which is expected to happen in a finite amount of time if the branching ratio $\eta<1$.\\

\subsection{Forecast evaluation}
    \label{sec:evaluation}
    The performance of each model is evaluated by calculating the log-likelihood of the testing data given the forecast.
    Specifically, we calculate the log-likelihood of $N_i$ earthquakes to occur in each bin $b_i$ of a spatial grid of $0.1 \degree$ latitude $\times$ $0.1\degree$ longitude.
    Here, $N_i$ is the number of earthquakes that actually occurred during the testing period in spatial bin $b_i$.
        
    The log-likelihood for $b_i$ is calculated based on the smoothed estimate of the probability of $N_i$ earthquakes to occur in $b_i$, where the probability estimate is based on the 100,000 simulations of the model in question.
    For smoothing we use Gaussian kernels with adaptive bandwidth 
    as described by \textcite{nandan2019forecastingfull}, with a fixed value of $\Omega=3.0$.
    To avoid arbitrary likelihood values due to extrapolation, we define a water-level likelihood for event counts larger than the maximum simulated event count in the respective bin.
    This waterlevel probability is defined as a uniform value of $100,001^{-1} / n_{extr}$, where $n_{extr}$ is the number of event counts larger than the maximum observed and smaller than a generously high maximum possible event count.
    Symbolically, this suggests that all other possible event counts could have been simulated in the $100,001^{st}$ simulation.
    Inevitably, the probabilities for non-extrapolated event counts are proportionally reduced such that the probabilities of all possible event counts add up to $1$.

    The total log-likelihood of the testing data is then given by the sum of log-likelihoods over all bins $b_i$.\\
    
    Two competing models can be compared by calculating the information gain (IG) of the alternative model $M_{alt}$ over the null model $M_0$, which is simply the difference in log-likelihood of observing the testing data.
    The mean information gain (MIG) is calculated as the mean over all testing periods. 
    We accept the superiority of $M_{alt}$ over $M_0$ when we reject the null hypothesis that $M_{alt}$ does not outperform $M_0$. 
    To decide whether to reject the null hypothesis, we perform a one-sided t-test on the set of IGs for all testing periods.
    This assumes that the IG values follow a normal distribution, and we test whether or not the MIG of $M_{alt}$ versus $M_0$ is significantly positive.
    We reject the null hypothesis when a $p$-value of less than 0.05 is observed.\\
    
    The spatially and temporally homogeneous Poisson process (STHPP) model forecasts the same number of events in all spatial bins.
    This number of events forecasted, $N_{fc}$, is given by
    
    \begin{equation}
        N_{fc} = \frac{N_{train}\cdot T_{test}}{T_{train} \cdot N_{bins}},
    \end{equation}
    where $n_{train}$ is the number of events observed in the training period, $T_{train}$ is the length of the training period in days,  $T_{test} = 30$ is the testing period length, and $N_{bins}$ is the total number of spatial bins.
    The log-likelihood for the STHPP model is calculated assuming a Poisson distribution of event numbers with mean $N_{fc}$ in each spatial bin.

\newpage
\printbibliography[heading=bibintoc]
\clearpage


%
%
%
%

\begin{figure}
    \centering
    \includegraphics[width = \textwidth]{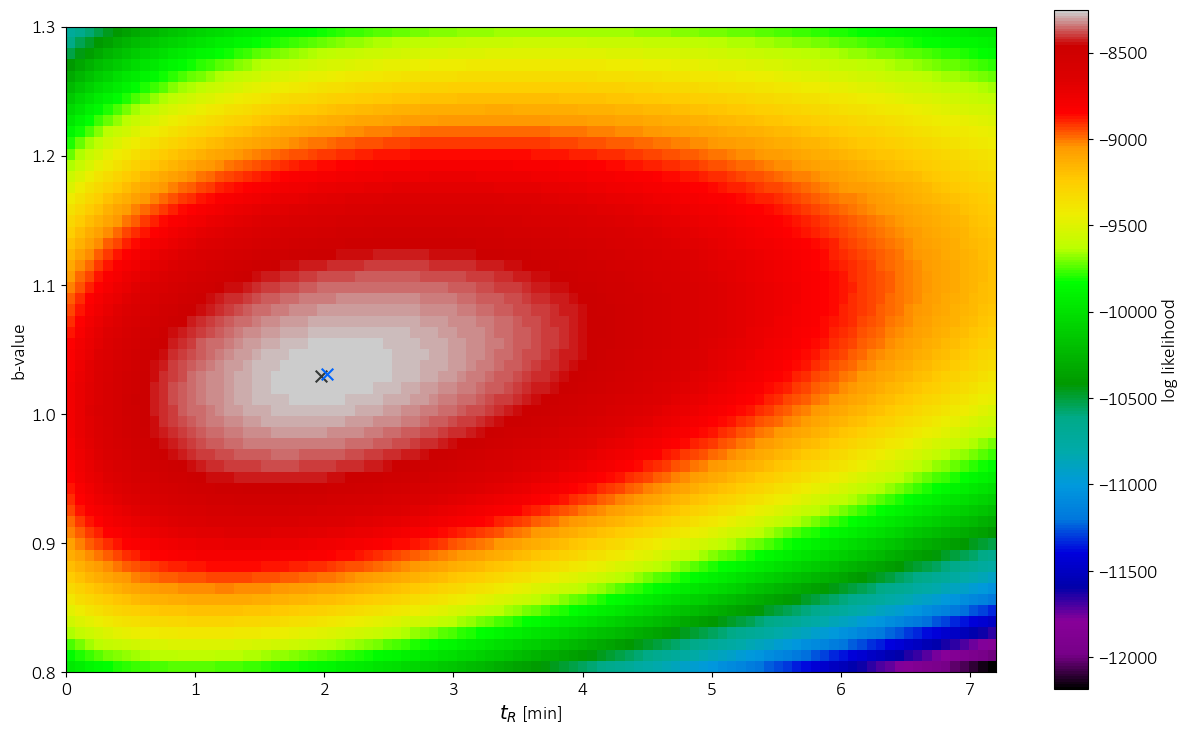}
    \caption{Log likelihood of observing the test data for different values of $t_R$ and $b$-value, when current rate is known. Black cross indicates true values used in simulation, blue cross indicates maximum likelihood estimators obtained.}
    \label{fig:tr_estimate}
\end{figure}